\documentclass[apjl]{emulateapj}
\usepackage{psfig,amsfonts,amsmath,graphicx,natbib,apjfonts,lscape}

\def\sw{Sw\,1644+57}

\def\cfa{1}
\def\ami{2}
\def\hu{3}
\def\mpifr{4}
\def\nrao{5}
\def\york{6}
\def\hrao{7}

\begin{document}

\title{Radio Monitoring of the Tidal Disruption Event
Swift\,J164449.3+573451. I. Jet Energetics and the Pristine
Parsec-Scale Environment of a Supermassive Black Hole}

\author{ 
E.~Berger\altaffilmark{\cfa},
A.~Zauderer\altaffilmark{\cfa},
G.~G.~Pooley\altaffilmark{\ami},
A.~M.~Soderberg\altaffilmark{\cfa},
R.~Sari\altaffilmark{\cfa,}\altaffilmark{\hu},
A.~Brunthaler\altaffilmark{\mpifr,}\altaffilmark{\nrao},
and M.~F.~Bietenholz\altaffilmark{\york,}\altaffilmark{\hrao}
}

\altaffiltext{\cfa}{Harvard-Smithsonian Center for Astrophysics, 60
Garden Street, Cambridge, MA 02138}

\altaffiltext{\ami}{Mullard Radio Observatory, Cavendish Laboratory,
Cambridge, CB3 0HE UK}

\altaffiltext{\hu}{Racah Institute of Physics, The Hebrew University,
91904 Jerusalem, Israel}

\altaffiltext{\mpifr}{Max-Planck-Institut f$\ddot{\rm u}$r
Radioastronomie, Auf dem H$\ddot{u}$gel 69, 53121 Bonn, Germany}

\altaffiltext{\nrao}{National Radio Astronomy Observatory, P.O. Box 0,
Socorro, NM 87801}

\altaffiltext{\york}{Department of Physics and Astronomy, York
University, Toronto, Ontario, Canada}

\altaffiltext{\hrao}{Hartebeesthoek Radio Astronomy Observatory, PO
Box 443, Krugersdorp, 1740 South Africa}

\begin{abstract} We present continued radio observations of the tidal
disruption event Swift\,J164449.3+573451 extending to $\delta t\approx
216$ days after discovery.  The data were obtained with the EVLA, AMI
Large Array, CARMA, the SMA, and the VLBA+Effelsberg as part of a
long-term program to monitor the expansion and energy scale of the
relativistic outflow, and to trace the parsec-scale environment around
a previously-dormant supermassive black hole (SMBH).  The new
observations reveal a significant change in the radio evolution
starting at $\delta t\approx 1$ month, with a brightening at all
frequencies that requires an increase in the energy by about an order
of magnitude, and an overall density profile around the SMBH of
$\rho\propto r^{-3/2}$ ($0.1-1.2$ pc) with a significant flattening at
$r\approx 0.4-0.6$ pc.  The increase in energy cannot be explained
with continuous injection from an $L\propto t^{-5/3}$ tail, which is
observed in the X-rays.  Instead, we conclude that the relativistic
jet was launched with a wide range of Lorentz factors, obeying
$E(>\Gamma_j)\propto \Gamma_j^{-2.5}$.  The similar ratio of duration
to dynamical timescale for \sw\ and GRBs suggests that this result may
be applicable to GRB jets as well.  The radial density profile may be
indicative of Bondi accretion, with the inferred flattening at $r\sim
0.5$ pc in good agreement with the Bondi radius for a $\sim {\rm
few}\times 10^6$ M$_\odot$ black hole.  The density at $\sim 0.5$ pc
is about a factor of 30 times lower than inferred for the Milky Way
galactic center, potentially due to a smaller number of mass-shedding
massive stars.  From our latest observations ($\delta t\approx 216$ d)
we find that the jet energy is $E_{\rm j,iso}\approx 5\times 10^{53}$
erg ($E_j\approx 2.4\times 10^{51}$ erg for $\theta_j=0.1$), the
radius is $r\approx 1.2$ pc, the Lorentz factor is $\Gamma_j\approx
2.2$, the ambient density is $n\approx 0.2$ cm$^{-3}$, and the
projected angular size is $r_{\rm proj}\approx 25$ $\mu$as, below the
resolution of the VLBA+Effelsberg.  Assuming no future changes in the
observed evolution and a final integrated total energy of $E_j\approx
10^{52}$ erg, we predict that the radio emission from \sw\ should be
detectable with the EVLA for several decades, and will be resolvable
with VLBI in a few years.  \end{abstract}


\section{Introduction}
\label{sec:into}

The discovery of the unusual $\gamma$-ray/X-ray transient
Swift\,J164449.3+573451 (hereafter, \sw), which coincided with the
nucleus of an inactive galaxy at $z=0.354$, has opened a new window
into high-energy transient phenomena, with potential implications to
our understanding of relativistic outflows in systems such as
gamma-ray bursts (GRBs) and active galactic nuclei (AGN).  The
prevailing interpretation for this event is the tidal disruption of a
star by a dormant supermassive black hole (SMBH) with a mass of
$M_{\rm BH}\sim 10^6-10^7$ M$_\odot$
(\citealt{bgm+11,bkg+11,ltc+11,zbs+11}; but see
\citealt{kp11,osj11,qk11} for alternative explanations).  The argument
for a tidal disruption origin is based on: (i) a positional
coincidence ($\lesssim 0.2$ kpc) with the host galaxy nucleus; (ii)
rapid time variability in $\gamma$-rays and X-rays ($\lesssim 10^2$
s), which requires a compact source of $\lesssim 0.15$ AU, a few times
the Schwarzschild radius of a $\sim 10^6$ M$_\odot$ black hole; (iii)
high $\gamma$-ray and X-ray luminosity of $\sim 10^{47}$ erg s$^{-1}$,
which exceeds the Eddington limit of a $\sim 10^6$ M$_\odot$ black
hole by $2-3$ orders of magnitude; (iv) a lack of previous radio to
$\gamma$-ray activity from this source to much deeper limits than the
observed outburst, pointing to a rapid onset; and (v) long-term X-ray
luminosity evolution following $L_X\propto t^{-5/3}$, as expected from
the fallback of tidally disrupted material (e.g.,
\citealt{ree88,sq09}).

Equally important, \sw\ was accompanied by bright radio synchrotron
emission, with an initial peak in the millimeter band ($F_\nu\approx
35$ mJy) and a steep spectral slope at lower frequencies indicative of
self-absorption (\citealt{zbs+11}; hereafter, ZBS11).  The properties
of the radio emission established the existence of a relativistic
outflow with a Lorentz factor of $\Gamma\sim {\rm few}$ (ZBS11,
\citealt{bgm+11}).  The spectral energy distribution also demonstrated
that the lack of detected optical variability required significant
rest-frame extinction ($A_V\gtrsim 5$ mag; ZBS11, \citealt{ltc+11}),
and that the X-rays were produced by a distinct emission component,
rather than inverse Compton scattering by the radio-emitting
relativistic electrons (ZBS11).  Finally, the evolution of the radio
emission on a timescale of $\delta t\sim 5-22$ d pointed to an ambient
density with a radial profile of roughly $\rho\propto r^{-2}$, as well
as a mild increase in the energy of the outflow (ZBS11).

The formation of a relativistic jet with dominant X-ray and radio
emission were not predicted in standard tidal disruption models (e.g.,
\citealt{ree88,sq09}), which instead focused on the thermal optical/UV
emission from the long-term accretion of the stellar debris.  A
signature of the latter process is a mass accretion rate that evolves
as $\dot{M}\propto t^{-5/3}$, presumably leading to emission with the
same temporal dependence (e.g., \citealt{kg99,gbm+08,vfg+11}).
Shortly before the discovery of \sw, \citet{gm11} investigated the
potential signature of a putative relativistic outflow, and concluded
that the interaction of the outflow with the ambient medium will lead
to radio emission on a timescale of $\delta t\sim 1$ yr (for typical
off-axis observers).  While the mechanism for the radio emission from
\sw\ is interaction with an external medium, the actual light curves
differ from the off-axis prediction.  To address this issue, in a
follow-up paper \citet{mgm11} (hereafter, MGM11) reconsidered the
model for a relativistic jet interacting with an ambient medium.  They
draw on the inferences from the early radio emission described in
ZBS11 to infer the properties of the environment and the jet kinetic
energy, and use this information to predict the future evolution of
the radio emission.

This long-term radio evolution is of great interest because it can
provide several critical insights:
\begin{itemize} 
\item The integrated energy release in the relativistic outflow,
including the anticipated injection from on-going accretion.
\item The density profile around a previously-dormant SMBH on $\sim
0.1-10$ pc scales, which cannot be otherwise probed in AGN.
\item The potential to spatially resolve the outflow with very long
baseline interferometry (VLBI), and hence to measure the dynamical
evolution (expansion and potentially spreading) of a relativistic jet.
\item Predictions for the radio emission from tidal disruption jets as
viewed by off-axis observers on timescales of months to years to
decades.
\end{itemize}
The energy scale and jet dynamics are of particular importance since
the total energy input and the structure of the jet may also have
implications for relativistic jets in GRBs and AGN.  The ability to
trace the environment on parsec scales provides a unique probe of gas
inflow or outflow around an inactive SMBH on scales that cannot be
probed outside of the Milky Way.  Finally, the long-term radio
emission from \sw\ will inform future radio searches for tidal
disruption events (TDEs) that can overcome the low detection rate in
$\gamma$-rays/X-rays (due to beaming), and obscuration due to
extinction in the optical/UV (as in the case of \sw).

To extract these critical properties we are undertaking long-term
monitoring of the radio emission from \sw\ using a wide range of
centimeter- and millimeter-band facilities.  Here we present radio
observations of \sw\ that extend to $\delta t\approx 216$ d, and use
these observations to determine the evolution of the total energy and
ambient density.  We find that the evolution of both quantities
deviates from the behavior at $\delta t\lesssim 1$ month (presented in
ZBS11), thereby providing crucial insight into the structure of the
relativistic outflow and the ambient medium.  This paper is the first
in a series that will investigate the long-term radio evolution of
\sw\ and the implications for relativistic jets and parsec-scale
environments around supermassive black holes, including efforts to
resolve the source with VLBI and to measure polarization.

The current paper is organized as follows.  We describe the radio
observations in \S\ref{sec:obs}, and summarize the radio evolution at
$\delta t\approx 5-216$ d in \S\ref{sec:prem}.  In \S\ref{sec:model}
we present our modeling of the radio emission, which utilizes the
formulation of MGM11.  The implications for the energy scale and
ambient density are discussed in \S\ref{sec:energy} and
\S\ref{sec:density}, respectively, and we finally consider the
implications for relativistic jets and the parsec-scale environments
of SMBHs in \S\ref{sec:implic}.

\section{Radio Observations}
\label{sec:obs}

Although \sw\ first triggered the {\it Swift} Burst Alert Telescope on
2011 March 28.55 UT, discernible $\gamma$-ray emission was detected
starting on 2011 March 25 UT \citep{bkg+11}.  We therefore consider
2011 March 25.5 UT to be the actual initial time for the event and the
associated relativistic outflow that powers the observed radio
emission.  Our radio observations of \sw\ commenced on 2011 March
29.36 UT ($19.4$ hr after the {\it Swift} trigger and about $3.9$ d
after the initial $\gamma$-ray detection).  Observations extending to
$\delta t\approx 26$ d were presented in ZBS11.  Here we report
observations extending to $\delta t\approx 216$ d.  Throughout the
paper we use the standard cosmological constants with $H_0=70$ km
s$^{-1}$ Mpc$^{-1}$, $\Omega_m=0.27$ and $\Omega_\Lambda=0.73$.

\subsection{Expanded Very Large Array}
\label{sec:evla}

We observed \sw\ with the EVLA\footnotemark\footnotetext{The EVLA is
operated by the National Radio Astronomy Observatory, a facility of
the National Science Foundation operated under cooperative agreement
by Associated Universities, Inc.  The observations were obtained as
part of programs 10C-145, 11A-262, and 11A-266} using the new Wideband
Interferometric Digital Architecture (WIDAR) correlator to obtain up
to 2 GHz of bandwidth at several frequencies.  At all frequencies, we
used 3C286 for bandpass and flux calibration.  At 1.4 GHz, we used
J1634+6245 for phase calibration.  For phase calibration at all other
frequencies, we used J1638+5720, and also included a third calibrator,
J1639+5357, at 5.8 GHz.  The data were reduced and imaged with the
Astronomical Image Processing System (AIPS) software package.  The
observations are summarized in Table~\ref{tab:data}.  Minor changes to
data values with respect to those reported in ZBS11 are due to
additional flagging of the data.  The errors reported in
Table~\ref{tab:data} are statistical uncertainties only; the overall
systematic uncertainty in the flux calibration is $\lesssim 5\%$.

\subsection{AMI Large Array}
\label{sec:ami}

We observed with the AMI Large Array at 15.4 GHz with a bandwidth of
3.75 GHz.  The maximum baseline is about 110 m, with a resulting
angular resolution of 25 arcsec.  Observations ranged in duration from
45 min to 11 hr.  Observations of the compact source J1638+5720 were
interleaved at intervals of 10 min as a phase reference, and the flux
density scale was established by regular observations of the
calibrators 3C48 and 3C286.  The telescope measures linearly-polarized
signals (Stokes I+Q).  The observations are summarized in
Table~\ref{tab:data}.

\subsection{Combined Array for Research in Millimeter Astronomy}
\label{sec:carma}

We observed \sw\ with CARMA at frequencies of 87.3 and 93.6 GHz with a
total bandwidth ranging between 6.8 and 7.8 GHz.  We used Neptune as
our primary flux calibrator, and J1824+568 and J1638+573 as bandpass
and phase calibrators.  The overall uncertainty in the absolute flux
calibration is $\approx 20\%$.  Data calibration and imaging were done
with the MIRIAD and AIPS software packages.  The observations are
summarized in Table~\ref{tab:data}.  We note some changes with respect
to the flux densities reported in ZBS11, especially those epochs where
the total integration time on source was less than $\sim 20$ min
($\delta t\approx 10-25$ d).

\subsection{Submillimeter Array}
\label{sec:sma}

We observed \sw\ with the SMA using at least seven of the eight
antennas, in a wide range of weather conditions, with $\tau_{\rm 225}$
ranging from 0.04 to 0.3.  In each observation we combined the two
sidebands, each with a bandwidth of 4 GHz separated by 10 GHz, to
increase the signal-to-noise ratio.  The data were calibrated using
the MIR software package developed at Caltech and modified for the
SMA, while for imaging and analysis we used MIRIAD.  Gain calibration
was performed using J1642+689, 3C345, and J1849+670.  Absolute flux
calibration was performed using real-time measurements of the system
temperatures, with observations of Neptune to set the overall scale.
Bandpass calibration was done using 3C454.3, J1924-292, and 3C279.
The observations are summarized in Table~\ref{tab:data}.

\subsection{Very Long Baseline Interferometry}
\label{sec:vlba}

We observed \sw\ with the NRAO Very Long Baseline Array
(VLBA\footnotemark\footnotetext{The data were collected as part of
programs BS210 and BS212.}) and the 100-m Effelsberg telescope in six
epochs\footnotemark\footnotetext{The dates of the observations are:
2011 April 2, April 8, May 5, May 28, July 17, and September 17 UT.}
between 2011 April 2 and September 17 UT at 8.4 and 22 GHz.  The
observations were performed with eight frequency bands of 8 MHz
bandwidth each in dual circular polarization, resulting in a total
data rate of 512 Mbps.  ICRF\,J1638+5720 \citep{Ma_etal_1998}, located
only 0.92$^\circ$ from \sw\ was used for phase-referencing at both
frequencies.  At 22 GHz, we switched between the target and calibrator
every 40 s, while at 8.4 GHz we spent 40 s on the calibrator and 90 s
on the target.  This resulted in a total integration time on \sw\ of
about 93 min at 8.4 GHz and 78 min at 22 GHz in epochs 1, 2, 5, and 6.
Epochs 3 and 4 were 1 hour shorter, resulting in integration times on
source of 75 min at 8.4 GHz and 60 min at 22 GHz.  A second
calibrator, J1657+5705 \citep{Beasley_etal_2002}, was observed for 6
min at each frequency to check the overall data calibration.  At 22
GHz, we also performed $\sim 30$-min blocks of geodetic observations
to perform atmospheric calibration (for details see
\citealt{BrunthalerRF2005,Reid_etal_2009}).

The results from epoch 1 were presented in ZBS11.  Here, we only
discuss the 22 GHz data; the full VLBI data set will be presented
together with future VLBI observations in an upcoming paper.  The data
were correlated at the VLBA Array Operations Center in Socorro, New
Mexico and calibrated using AIPS and ParselTongue
\citep{Kettenis_etal_2006}.  We applied the latest values of the Earth
orientation parameters and performed zenith delay corrections based on
the results of the geodetic block observations.  Total electron
content maps of the ionosphere were used to correct for ionospheric
phase changes.  Amplitude calibration used system temperature
measurements and standard gain curves.  We performed a ``manual
phase-calibration'' using the data from one scan of J1638+5720 to
remove instrumental phase offsets among the frequency bands.  We then
fringe-fitted the data from J1638+5720.  Since J1638+5720 has extended
structure, we performed phase self-calibration, and later amplitude
and phase self-calibration on J1638+5720 to construct robust models of
J1638+5720 at both frequencies.  These models were then used to
fringe-fit the data again.  Finally, the calibration was transferred
to \sw\ and J1657+5705.  The data were imaged in AIPS using robust
weighting (with {\tt ROBUST=0}).

A linear fit to the positions of \sw\ from epochs 2--6 gives a proper
motion of $81\pm 37$ $\mu$as yr$^{-1}$ in right ascension and $46\pm
59$ $\mu$as yr$^{-1}$ in declination, with an additional systematic
uncertainty of about 40 $\mu$as yr$^{-1}$.  This is consistent with no
detectable motion.  The corresponding $3\sigma$ upper limit on the
motion at the redshift of \sw\ is $\lesssim 1$ pc.  Since the first
epoch was correlated at a different position, it was not included in
this fit to avoid second order systematic errors.

To obtain an accurate estimate of the angular size in our latest epoch
(2011 September 17 UT), we fit Gaussian models directly to the
visibility data.  We first self-calibrated the data in phase using
solution intervals of 20 min, and then fitted a circular Gaussian
model using a weighted least-squares fit.  The source is not resolved
in our observation.  For unresolved sources the fitted size can be
strongly correlated with the antenna amplitude gains, which are only
imprecisely known.  Our uncertainty on the fitted FWHM size, and
consequently our upper limit, was estimated from a Monte-Carlo
simulation in which we randomly varied the amplitude gains of the
antennas by conservative factors of $\pm 25$\% (for a more elaborate
discussion of the uncertainties in estimating angular source sizes
from a similar procedure see \citealt{Bietenholz_etal_2010}).  The
$3\sigma$ upper limit on the FWHM source size is $0.22$ mas,
corresponding to a projected physical diameter of $\lesssim 1.1$ pc.
For spherical expansion this corresponds to an upper limit on the
apparent expansion velocity of $\lesssim 3.8c$.  For a collimated
relativistic source with an opening angle $\theta_j$ the projected
diameter is $\approx 4\Gamma^2ct\theta_j$, indicating that for
$\theta_j=0.1$ the limit on the size of \sw\ corresponds to
$\Gamma\lesssim 5$ at $\delta t\approx 176$ d.  We compare this result
with estimates of the source size from modeling of the radio emission
in \S\ref{sec:energy}.

\section{Observed Evolution of the Radio Emission}
\label{sec:prem}

\begin{figure}
\epsscale{1}
\plotone{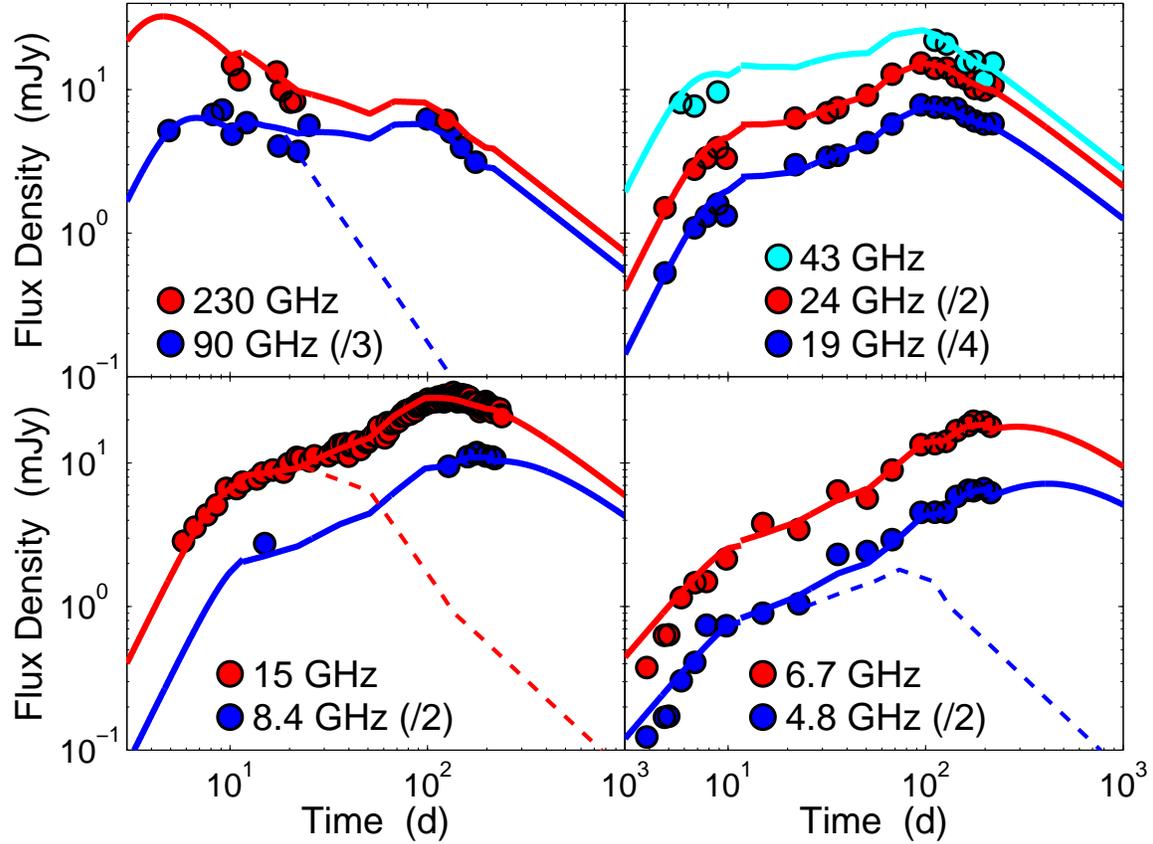}
\caption{Radio light curves of \sw\ extending to $\delta t\approx 216$
d.  The data at $\delta t\approx 5-22$ d were previously presented in
ZBS11.  The solid lines are models based on independent fits of
broad-band SEDs (Figure~\ref{fig:specs}) using the model described in
\S\ref{sec:model} (see also MGM11).  The dashed lines are the
predicted light curves from MGM11, which assumed a constant energy and
a steady density profile of $\rho\propto r^{-2}$.  The secondary
maximum in the millimeter band and the continued increase in
brightness to a peak time of $\delta t\gtrsim 100-200$ d in the
centimeter bands require an increase in the energy and density
relative to the initial evolution.
\label{fig:lcs}} 
\end{figure}

The light curves at observed frequencies of 4.9, 6.7, 15.4, 19, 24,
43, 90, and 230 GHz are presented in Figure~\ref{fig:lcs}.  As
described in ZBS11, the light curves at 4.9 to 24 GHz exhibit an
initial rapid increase, with a flux density of $F_\nu\propto
t^{1.5}-t^2$, followed by a shallower increase, with $F_\nu\propto
t^{0.5}$.  In the millimeter band, the light curves peak on a
timescale of a few days and then decline as $F_\nu\propto t^{-1}$.
This evolution is due to a synchrotron spectrum that is optically thin
in the millimeter band and optically thick in the centimeter band,
with a peak frequency and flux density that decline as a function of
time as $\nu_p\propto t^{-1.3}$ and $F_{\nu,p}\propto t^{-0.8}$
(ZBS11).  The properties of the spectral energy distribution, along
with measurements of interstellar scintillation, established that the
outflow is relativistic ($\Gamma\approx 2-5$), that the density
profile of the circumnuclear medium (CNM) is roughly $\rho_{\rm
CNM}\propto r^{-2}$, and that the post-shock energy increases as
$E\propto t^{0.5}$ (ZBS11).

The most striking result from our new observations is that the
millimeter flux on a timescale of $\sim 100$ d is significantly
brighter than expected from an extrapolation of the early decline.
Indeed, at 90 GHz the brightness is comparable to the initial peak at
$\delta t\sim 10$ d.  A similar effect is observed at lower
frequencies, where the light curves exhibit an upturn starting at
$\approx 30$ d, with $F_\nu\propto t^{0.7}$.  The 90 GHz light curve
is already in a declining (i.e., optically thin) phase at $\delta
t\gtrsim 100$ d, while the light curves at 4.9 to 24 GHz peak at
$\delta t\approx 100$ d (19 and 24 GHz), $\delta t\approx 125$ d (15.4
GHz), $\delta t\approx 170$ d (8.4 GHz), and $\delta t\approx 190$ d
(4.9 and 6.7 GHz), with peak flux densities of $\approx 31$ mJy,
$\approx 28$ mJy, $\approx 23$ mJy, and $\approx 20$ mJy,
respectively.  These observations are at odds with the light curve
evolution predicted by MGM11 based on our radio data at $5-22$ d,
using a model that assumes a constant energy and a steady $\rho\propto
r^{-2}$ CNM profile (Figure~\ref{fig:lcs}).

The observed change in the light curve evolution at $\delta t\gtrsim
1$ month requires an increase in the outflow kinetic energy and/or the
CNM density.  The evolution of the peak frequency, roughly
$\nu_p\propto t^{-1}$ at $\delta t\sim 30-100$ d, is slower than
expected for the characteristic synchrotron frequency, $\nu_m\propto
t^{-1.5}$ (see \S\ref{sec:model}).  Since $\nu_m$ only depends on the
kinetic energy, $\nu_m\propto E_j^{0.5}t^{-3/2}$ \citep{gs02}, the
observed evolution requires an increase in the energy, with
$E_j\propto t$.  At the same time, the similar light curve evolution
at 4.9 and 24 GHz, which straddle the self-absorption frequency (see
\S\ref{sec:model}), indicates that the density must also change.  In
particular, at 4.9 GHz we expect $F_\nu\propto E_j^{1.5}\,A^{-1.5}\,
t^{0.5}$, while at 24 GHz we expect $F_\nu\propto E_j^{5/6}\,A^{0.5}\,
t^{-0.5}$; here $A$ is a fiducial density that parametrizes the CNM
density, with $\rho=Ar^{-2}$.  For the observed light curve evolution
of $F_\nu\propto t^{0.7}$ at both frequencies, and using the inferred
evolution of $E_j$, we find that $A\propto t^{0.85}$.  Thus, both the
energy and density normalization appear to increase at $\delta
t\gtrsim 30$ d.  In the next section we model the radio emission in
detail to extract the temporal evolution of the energy and density.

\section{Detailed Modeling of the Radio Emission}
\label{sec:model}

To determine the temporal evolution of the relativistic outflow and
the radial density profile we use the formulation of MGM11, which
draws on the GRB afterglow model of \citet{gs02} (hereafter, GS02).
The MGM11 model is largely motivated by the basic properties of the
outflow that were determined in our initial study of the radio
emission (ZBS11), as well as by the observed evolution of the X-ray
emission \citep{bkg+11}.  The model assumes that the outflow is
collimated, with an opening angle $\theta_j$, and has a Lorentz factor
of $\Gamma_j$; in ZBS11 we demonstrated that $\Gamma_j\sim {\rm few}$
(see also \citealt{bgm+11}).  The kinetic luminosity of the outflow,
$L_{\rm j,iso}$, is assumed to be constant for a timescale $t_j$,
followed by a decline as $L_{\rm j,iso}\propto t^{-5/3}$ at $t\gtrsim
t_j$.  This is expected for fallback accretion in a TDE, and is
observationally motivated by the evolution of the X-ray light curve,
which also indicates that $t_j\approx 10^6$ s \citep{bkg+11}.  As a
result of the long-term injection of energy, the expectation is that
the kinetic energy will gradually increase from an initial value of
$E_{\rm j,iso}=L_{\rm j,iso}\,t_j$ to a final level of $(5/2)\,L_{\rm
j,iso}\,t_j$.  From the initial X-ray luminosity, $L_{\rm X,iso}
\approx 6\times 10^{46}$ erg s$^{-1}$, the expected total energy is
$E_{\rm j,iso}\approx 3\times 10^{53}\,(\epsilon_X/0.5)^{-1}$ erg,
where $\epsilon_X$ is the efficiency of the jet in producing X-rays
(MGM11), while the predicted beaming-corrected energy is $E_j\approx
1.5\times 10^{51}$ erg.

The interaction of the outflow with the CNM leads to synchrotron
emission due to the acceleration of electrons and amplification of
magnetic fields.  Based on the early radio observations, ZBS11
demonstrated that the CNM density roughly follows $\rho_{\rm
CNM}\propto r^{-2}$ (hereafter, Wind medium), which is the profile
adopted for the analysis in MGM11.  The Lorentz factor of the fluid
behind the forward shock is $\Gamma_{\rm sh,FS}=\Gamma_j(1-\Gamma_j
\sqrt{2\,n_{\rm CNM}/7\,n_j})$, where $n_{\rm CNM}$ is the
circumnuclear density and $n_j=L_{\rm j,iso}/4\pi r^2m_pc^3\Gamma_j^2$
is the density of the ejecta (see MGM11); this relation is appropriate
for the case of a Newtonian reverse shock, i.e., $\Gamma_j^2\ll
n_j/n_{\rm CNM}$ \citep{sp95}, which as we find below is appropriate
for \sw\ (Table~\ref{tab:params}).  A key difference between the MGM11
model and the GS02 GRB afterglow model is that the former includes a
suppression of the flux density by a factor of $(\Gamma_{\rm
sh,FS}\theta_j)^2/2$ due to the finite extent of the collimated
outflow (i.e., $\Gamma_{\rm sh,FS} \lesssim 1/\theta_j$).  Thus, the
synchrotron emission from the relativistic outflow is determined by
$L_{\rm j,iso}$, $n_{\rm CNM}$, and the fractions of post-shock energy
in the radiating electrons ($\epsilon_e$) and the magnetic field
($\epsilon_B$).

We use this model (Equations 7--11 of MGM11), along with the smoothing
formulation of GS02, to fit snap-shot broad-band spectra of \sw\ on a
timescale of $\delta t\approx 5-216$ d.  In each epoch we fix
$\epsilon_e=\epsilon_B=0.1$ and $p=2.5$, and we then determine the
best-fit values of $L_{\rm j,iso}$ and $n_{\rm CNM}$; here $p$ is the
power law index of the electron Lorentz factor distribution,
$N(\gamma)\propto\gamma^{-p}$.  For our choice of $p$ value the
synchrotron frequencies (self-absorption: $\nu_a$; peak: $\nu_m$) and
flux normalization [$F_\nu(\nu=\nu_a)$] for $t\gtrsim t_j$ are given
by (MGM11, GS02):
\begin{eqnarray}
\nu_a(t)&=&4.2\times 10^9\, \epsilon_{e,-1}^{-1}\,
\epsilon_{B,-1}^{0.2}\, L_{\rm j,iso,48}^{-0.4}\, t_{j,6}^{-1}\,
n_{18}^{1.2}\, \left(\frac{t}{t_j}\right)^{-0.6}\,\, {\rm Hz},\,
\label{eqn:nua}\\
\nu_m(t)&=&3.5\times 10^{11}\, \epsilon_{e,-1}^{2}\,
\epsilon_{B,-1}^{0.5}\, L_{\rm j,iso,48}^{0.5}\, t_{j,6}^{-1}\,
\left(\frac{t}{t_j}\right)^{-1.5}\,\, {\rm Hz},\, \label{eqn:num}\\
F_\nu[\nu_a(t)]&=&345\, \epsilon_{e,-1}\, L_{\rm j,iso,48}^{1.5}\,
t_{j,6}^{2}\, n_{18}^{-1.5}\, \theta_{j,-1}^2\, \nu_{a,10}^2\,
\left(\frac{t}{t_j}\right)^{0.5}\,\, {\rm mJy},\,\label{eqn:fnua}
\end{eqnarray}
where we use the notation $X\equiv 10^{y}\,X_{y}$, and $n_{18}$ is the
CNM density ($n_{\rm CNM}$) at a fiducial radius of $10^{18}$ cm.  At
$t\lesssim t_j$ we use the same equations but with the time
dependencies modified to $\nu_a\propto t^{-1}$, $\nu_m\propto t^{-1}$,
and $F_\nu(\nu=\nu_a)\propto t^2$ (MGM11).  The synchrotron spectrum
is given by (c.f., Spectrum 1 of \citealt{gs02}):
\begin{eqnarray}
F_\nu=F_\nu(\nu_a)\left[\left(\frac{\nu}{\nu_a}\right)^{-s_1\beta_1}+
\left(\frac{\nu}{\nu_a}\right)^{-s_1\beta_2}\right]^{-1/s_1}\times
\nonumber \\ \left[1+\left(\frac{\nu}{\nu_m}\right)^{s_2(\beta_2-
\beta_3)}\right]^{-1/s_2},
\label{eqn:spec1}
\end{eqnarray} 
where $s_1$ and $s_2$ are smoothing parameters (GS02), and
$\beta_1=2$, $\beta_2=1/3$, and $\beta_3=(1-p)/2$ are the power law
indices for each segment of the synchrotron spectrum.

Equations~\ref{eqn:nua} and \ref{eqn:spec1} are appropriate when
$\nu_a<\nu_m$, and have to be modified when $\nu_m<\nu_a$ as follows
(GS02, MGM11\footnotemark\footnotetext{Note that MGM11 use the
scalings appropriate for Spectrum 3 of GS02, while we use the
appropriate Spectrum 2 with $\nu_m<\nu_a<\nu_c$.}):
\begin{equation}
\nu_a\propto t^{-3(p+2)/2(p+4)},
\end{equation}
\begin{eqnarray}
F_\nu=F_\nu(\nu_m)\left[\left(\frac{\nu}{\nu_m}\right)^{-s_1\beta_1}+
\left(\frac{\nu}{\nu_m}\right)^{-s_1\beta_2}\right]^{-1/s_1}\times
\nonumber \\ \left[1+\left(\frac{\nu}{\nu_a}\right)^{s_2(\beta_2-
\beta_3)}\right]^{-1/s_2},
\label{eqn:spec2}
\end{eqnarray}
where the spectrum is now normalized at $\nu=\nu_m$, $s_1$ and $s_2$,
take on different values (GS02) and $\beta_1=2$, $\beta_2=5/2$, and
$\beta_3=(1-p)/2$.  To smoothly connect the evolution in the two
phases we use a weighted average of Equations~\ref{eqn:spec1} and
\ref{eqn:spec2}, with the weighting determined by the time difference
relative to the transition time, $t_{\rm am}$, defined as
$\nu_a(t_{\rm am})=\nu_m(t_{\rm am})$.  We find that $t_{\rm am}
\approx 275$ d, so this transition does not affect the data presented
in this paper.

\begin{figure}
\epsscale{1}
\plotone{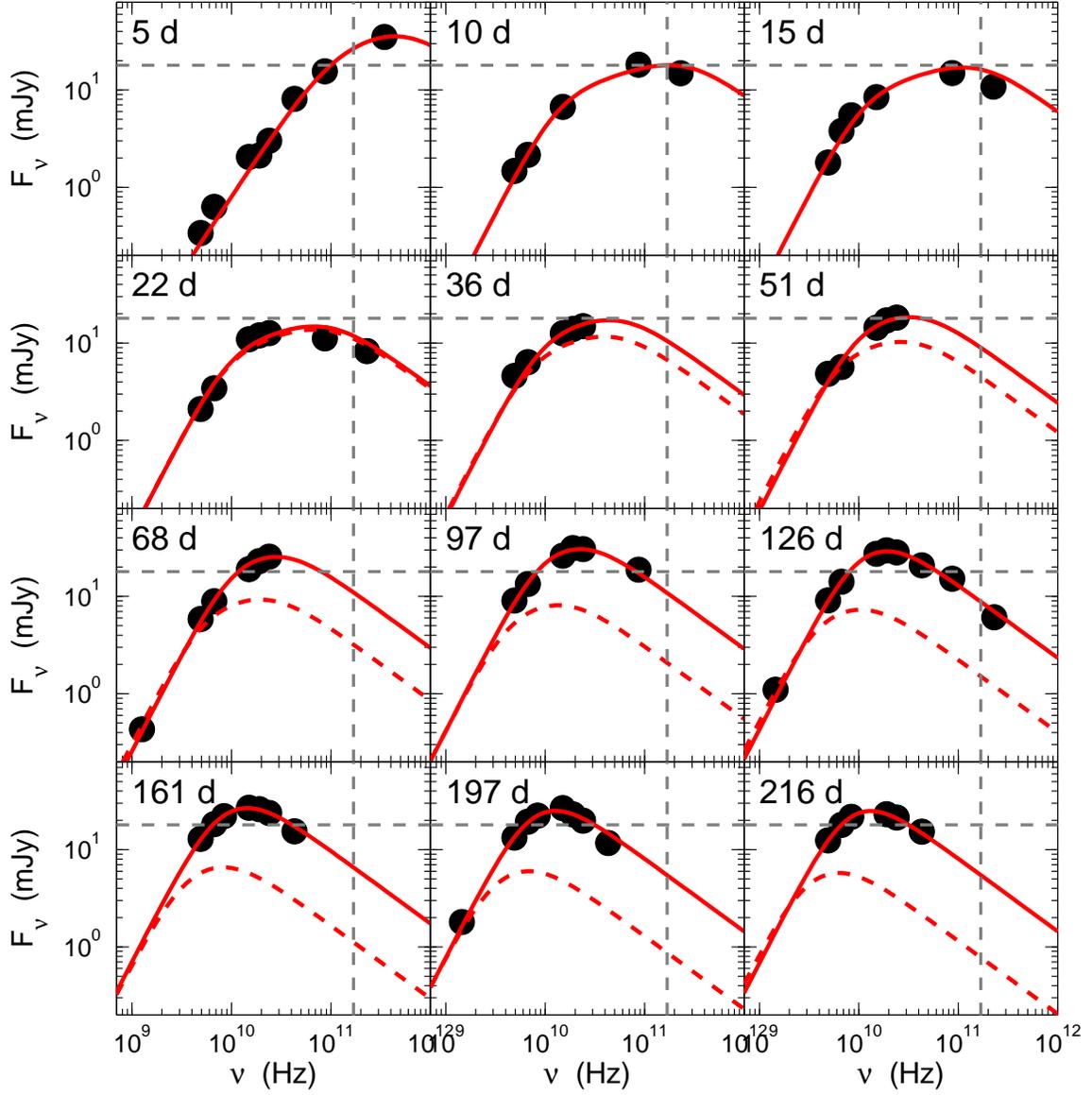}
\caption{Radio spectral energy distributions of \sw\ at $\delta
t\approx 5-216$ d.  The data at $\delta t\approx 5-22$ d were
previously presented in ZBS11.  The solid lines are fits based on the
model described in \S\ref{sec:model} (see also MGM11).  In each epoch
we fit for $L_{\rm j,iso}$ and $n_{18}$ with fixed values of
$\epsilon_e=\epsilon_B=0.1$ and $p=2.5$.  The dashed gray lines mark
the peak flux density and peak frequency at $\delta t=10$ d to help
track the evolution of the spectrum as a function of time.  The dashed
red lines mark the expected SEDs based on the evolution at $\delta
t\approx 5-22$ d.
\label{fig:specs}} 
\end{figure}

Instead of imposing a specific temporal evolution on $L_{\rm j,iso}$
and $n_{\rm CNM}$, we model each broad-band radio spectrum
independently to extract the evolution of these quantities, and in
turn the time evolution of the emission radius ($r$) and $\Gamma_{\rm
sh}$ (and hence $\Gamma_j$), as well as the radial density profile.
The results of the individual fits are shown in Figure~\ref{fig:specs}
and the relevant extracted parameters are plotted in
Figure~\ref{fig:params} and summarized in Table~\ref{tab:params}.  The
broad-band SEDs reveal a complex evolution.  At $\delta t\approx 5-22$
d both the peak frequency and peak flux density decrease with time (as
noted by ZBS11), but starting at $\delta t\approx 36$ d the peak flux
density begins to increase, reaching a maximum at $\delta t\approx
100$ d, and subsequently declining again.

The VLBI size limit of $r_{\rm proj}\lesssim 0.55$ pc at $\delta
t\approx 175$ d can be used to set an independent upper bound on the
ratio $L_{\rm j,iso,48}/n_{18}$.  In particular, on this timescale we
expect $r_{\rm proj}\approx 2.9\,t_{j,6}(L_{\rm j,iso,48}/
n_{18})^{0.5}\theta_j$ pc, indicating that $L_{\rm j,iso,48}/n_{18}
\lesssim 3.6$ (for $\theta_j=0.1$).  From the results of our radio
modeling we find on a similar timescale that $L_{\rm j,iso,48}\approx
0.45$ and $n_{18}\approx 3$ (Table~\ref{tab:params}), and hence
$L_{\rm j,iso,48}/n_{18}\approx 0.15$ is in agreement with the VLBI
size limit.  On the other hand, if we assume $\epsilon_e=0.01$ the
resulting value of $L_{\rm j,iso,48}$ increases by about a factor of
15, while $n_{18}$ decreases by about a factor of 2, leading to
$L_{\rm j,iso,48}/n_{18}\approx 4$, in violation of the VLBI size
limit\footnotemark\footnotetext{Changing to $\epsilon_B=0.01$ has
almost no effect since both the energy and density increase by about a
factor of 3 compared to $\epsilon_B=0.1$.}  We therefore conclude from
the VLBI results that $\epsilon_e\gtrsim 0.01$.

\subsection{The Evolution of $E_{\rm j,iso}$}
\label{sec:energy}

\begin{figure}
\epsscale{1}
\plotone{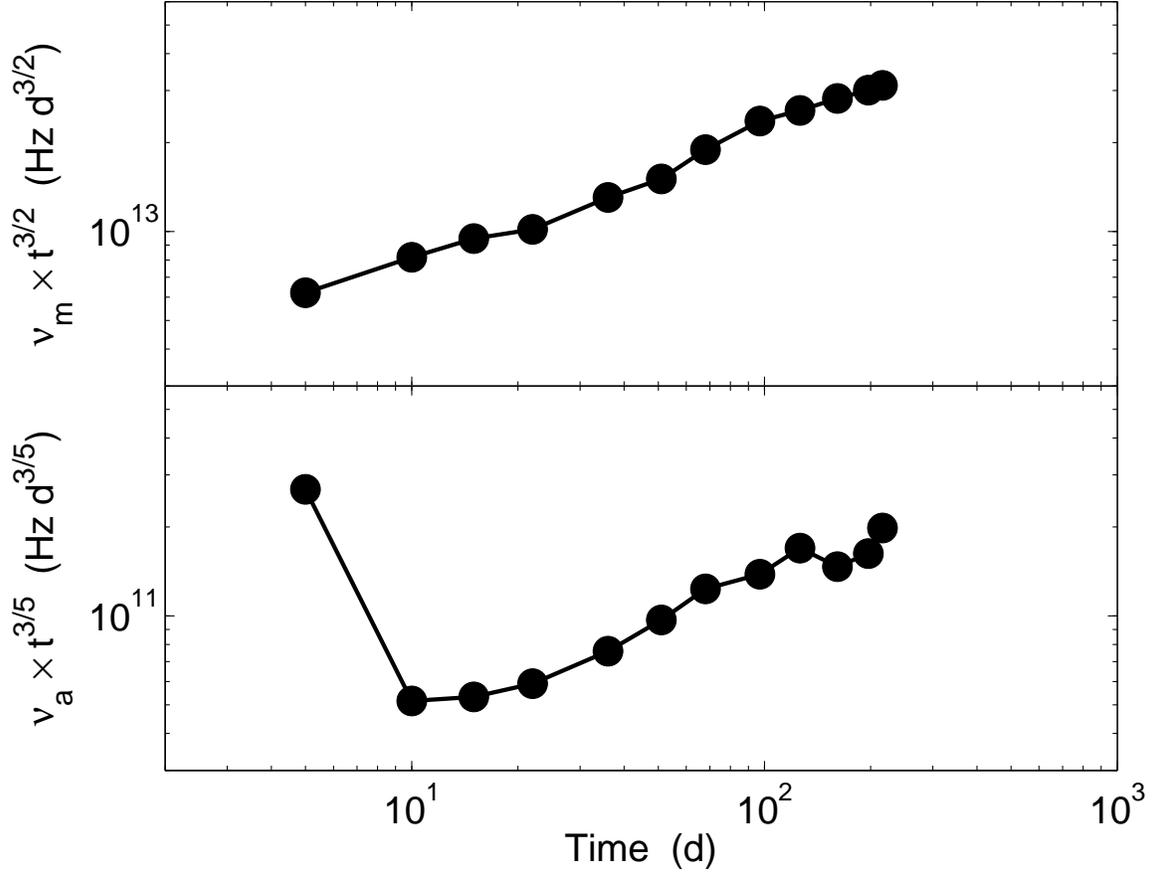}
\caption{Temporal evolution of the synchrotron frequencies $\nu_m$
(top) and $\nu_a$ (bottom) relative to the expected evolution in a
simple model with a constant energy and a Wind profile.  The shallower
decline of $\nu_m\propto E_j^{0.5}$, with a particularly shallow
evolution at $\delta t\approx 30-100$ d, is indicative of a continuous
increase in energy.  Similarly, the shallower decline of $\nu_a$,
followed by a rapid increase, is indicative of a density profile of
$\rho\propto r^{-1.5}$ and a flattening at $\delta t\approx 30-100$
d.
\label{fig:nuanum}} 
\end{figure}

A detailed analysis of the evolution of $\nu_m$ (shown in
Figure~\ref{fig:nuanum}) reveals a complex behavior.  While it
continuously declines as a function of time, the decline rate is
always shallower than the expected $t^{-3/2}$ for constant energy.
Since $\nu_m$ only depends on $L_{\rm j,iso}$ and the equipartition
fractions (which are not expected to change with time), the shallower
than expected evolution directly implies that $L_{\rm j,iso}$
continuously increases with time.  In particular, at $5-22$ d and
$\gtrsim 100$ d, the observed evolution of $\nu_m\propto t^{-1.15}$
indicates that $L_{\rm j,iso}\propto t^{0.7}$, while in the
intermediate phase ($22-100$) d, $\nu_m\propto t^{-0.95}$ points to a
more rapid increase in the energy scale, with $L_{\rm j,iso}\propto
t^{1.1}$.

\begin{figure}
\epsscale{1}
\plotone{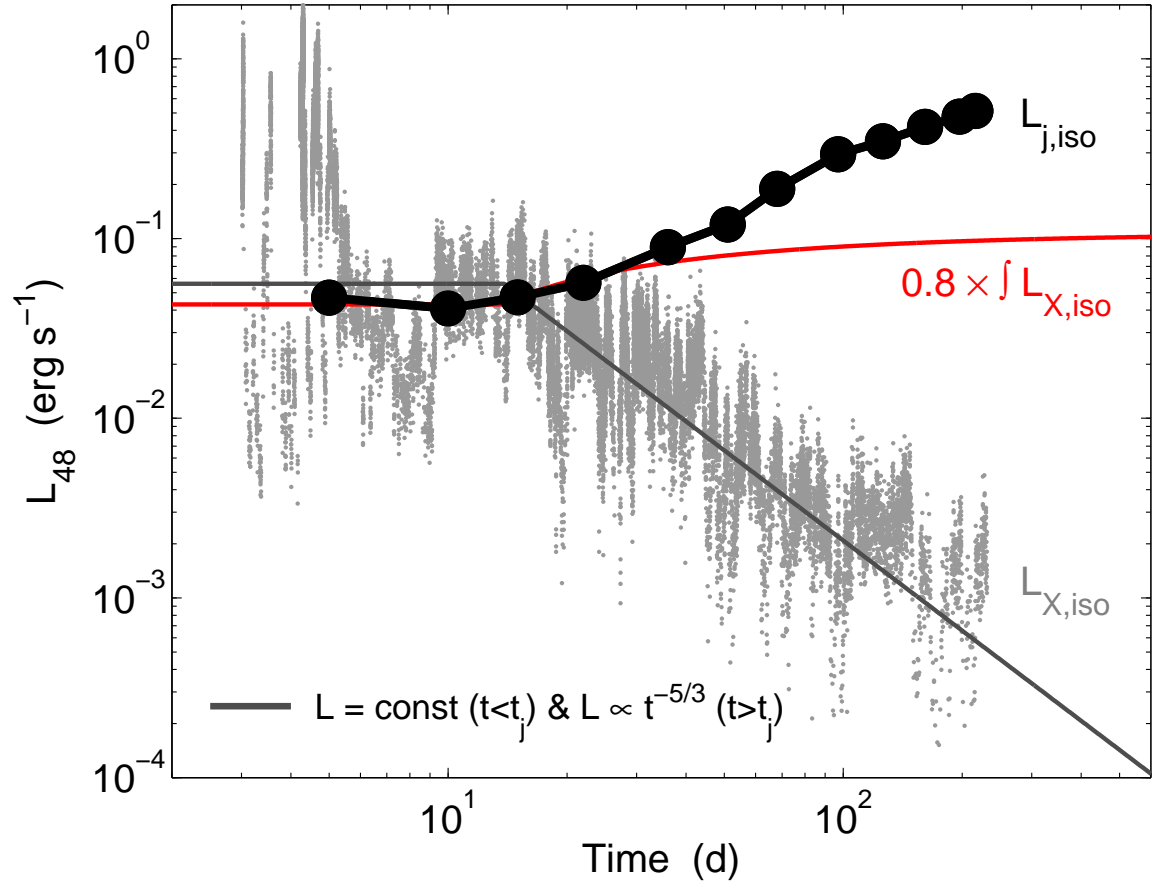}
\caption{Temporal evolution of the integrated luminosity (or
alternatively $E_{\rm j,iso}=L_{\rm j,iso}t_j$; black circles) in
comparison to the X-ray luminosity (gray dots) as parametrized with a
simple luminosity evolution (gray line).  The red curve is the
integrated luminosity derived from the simple model.  The observed
X-ray luminosity indicates that the fraction of total energy emitted
in X-rays is comparable to the energy in the relativistic outflow
(i.e., $\epsilon_X\approx 0.5$).  The large increase in energy
inferred from the radio observations cannot be explained by injection
from a $L\propto t^{-5/3}$ tail.
\label{fig:ljiso}} 
\end{figure}

The time evolution of $L_{\rm j,iso}$ is shown in
Figure~\ref{fig:ljiso}.  We find that at $\delta t\lesssim 20$ d
$L_{\rm j,iso}\approx 4.5\times 10^{46}$ erg s$^{-1}$ (or $E_{\rm
j,iso}\approx 4.5\times 10^{52}$ erg for $t_j=10^6$ s).  Changing the
values of the equipartition fractions from our assumed values
$\epsilon_e=\epsilon_B=0.1$, we find that $L_{\rm j,iso}$ increases by
about a factor of 3 if $\epsilon_B=0.01$ or by a factor of 15 if
$\epsilon_e=0.01$ ($\epsilon_e\lesssim 0.01$ is ruled out by the VLBI
limits on the source size).  These scalings hold for the overall time
evolution of $L_{\rm j,iso}$.

Also shown in Figure~\ref{fig:ljiso} is the expected evolution for a
model with $L_{\rm j,iso}={\rm const}$ at $t<t_j$ and $L_{\rm j,iso}
\propto t^{-5/3}$ at $t>t_j$, as inferred from the X-ray light curve
(\citealt{bkg+11}, MGM11).  In this model, we expect the integrated
value of $L_{\rm j,iso}$ (i.e., the total isotropic-equivalent energy)
to approach $(5/2)L_{\rm j,iso}$ as $t\rightarrow\infty$, and to
roughly double within $\delta t\sim 5t_j\approx 50-60$ d.  This seems
to be the case based on the radio data at $\delta t\lesssim 30$ d, but
the subsequent rapid increase in energy by about an order of magnitude
at $30-216$ d cannot be explained with continued injection from a
$t^{-5/3}$ tail.  On the other hand, the X-ray light curve of \sw\
(Figure~\ref{fig:ljiso}) agrees well with this simple luminosity
evolution, with $L_{\rm X,iso,0}\approx 6\times 10^{46}$ erg s$^{-1}$
at $t\lesssim 13$ d.  A comparison to the inferred value of $L_{\rm
j,iso,0}$ indicates a high efficiency\footnotemark\footnotetext{The
X-ray efficiency is instead $\epsilon_X\approx 0.3$ if
$\epsilon_B=0.01$ or $\approx 0.1$ if $\epsilon_e=0.01$.} of producing
X-rays of $\epsilon_X\approx 0.5$.

\begin{figure}
\epsscale{0.8}
\centering
\plotone{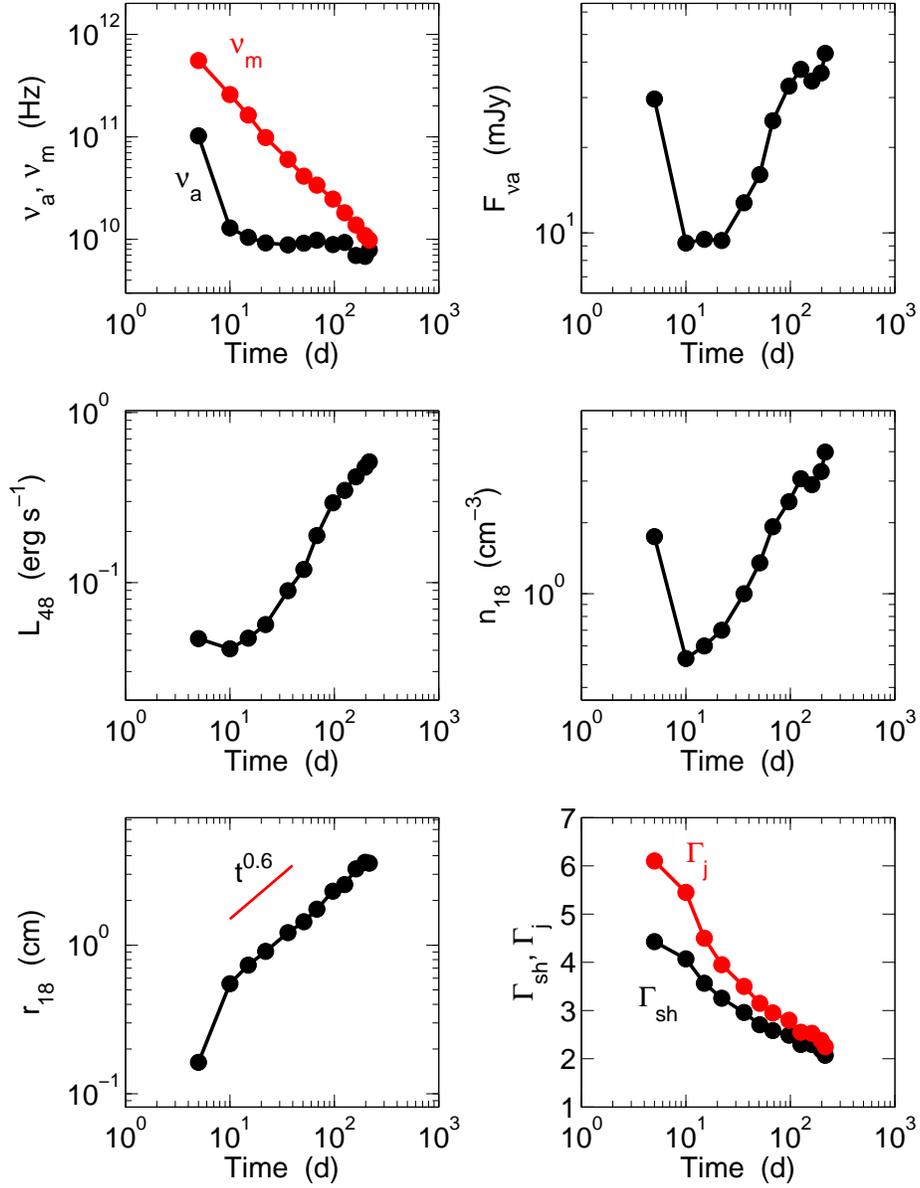}
\caption{Model and extracted parameters for each broad-band SED shown
in Figure~\ref{fig:specs}.  Shown are the time evolution of the
synchrotron parameters ($\nu_a$, $\nu_m$, and $F_{\nu_a}$), $L_{\rm
j,iso,48}$, $n_{18}$, $r$, $\Gamma_{\rm sh}$, and $\Gamma_j$.  The
substantial increase in energy and density is clearly seen.  In
addition, we find $r\propto t^{0.6}$, a steeper increase than expected
in a simple Wind model with constant energy ($r\propto t^{0.5}$).  
\label{fig:params}} 
\end{figure}

Since continuous energy injection from an $L\propto t^{-5/3}$ tail
cannot explain the inferred rise in integrated kinetic energy, the
radio data require a different energy injection mechanism.  One
possibility is that the outflow has a distribution of Lorentz factors,
with increasing energy as a function of decreasing Lorentz factor,
$E_j(>\Gamma_j)\propto\Gamma_j^{1-s}$ \citep{sm00}.  In this scenario
the energy will increase with time as material with a lower Lorentz
factor catches up with decelerated ejecta that initially had a higher
Lorentz factor.  The continuous injection of energy will also lead to
a more rapid increase in radius, with $r\propto t^{(1+s)/(7+s-2m)}$
\citep{sm00}; here $m$ is the power law index describing the CNM
density profile ($\rho\propto r^{-m}$).  In Figure~\ref{fig:params} we
plot the inferred radius as a function of time, and find that it
follows $r\propto t^{0.6}$, faster than expected in a simple Wind
model ($r\propto t^{0.5}$).  We also find from the density profile
that $m\approx 1.5$ (see \S\ref{sec:density}), thereby leading to
$s\approx 3.5$.  Thus, we expect in this scenario a fairly steep
profile for the energy as a function of Lorentz factor of
$E(>\Gamma_j)\propto \Gamma_j^{-2.5}$.

With this relation, and the inferred time evolution of the jet Lorentz
factor (Figure~\ref{fig:params}) we can infer the expected increase in
energy.  We find $\Gamma_j\approx 5.5$ at $\delta t=10$ d and
$\Gamma_j\approx 2.2$ at $\delta t=216$ d, indicating an expected
increase in energy by about a factor of 10 for our inferred profile.
This is in excellent agreement with the data.  Thus, a jet with a
distribution of Lorentz factors of $E(>\Gamma_j)\propto
\Gamma_j^{-2.5}$ naturally explains the evolution of source size and
the substantial increase in energy beyond the on-going input from a
$t^{-5/3}$ tail.

\subsection{The Radial Density Profile}
\label{sec:density}

\begin{figure}
\epsscale{1}
\plotone{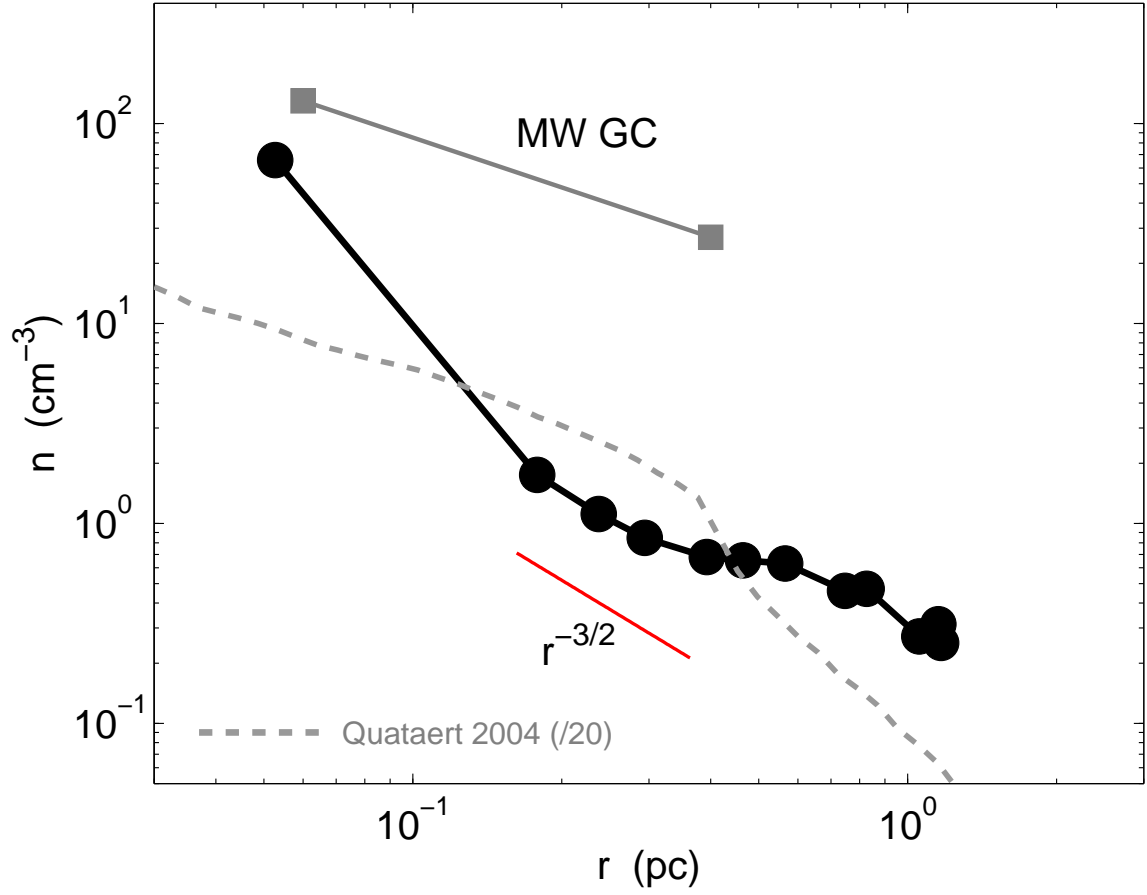}
\caption{Radial density profile in the inner parsec around \sw\ as
inferred from the radio observations (black circles).  The overall
profile follows $\rho\propto r^{-3/2}$, with a significant flattening
at $r\approx 0.4-0.6$ pc.  Following the flattening, the profile
appears to recover to $r^{-3/2}$ by about 1 pc.  Also shown is the
density inferred from X-ray observations of the Galactic center (gray
squares; \citealt{bmm+03}), which is about a factor of 30 times larger
at $\approx 0.5$ pc.  The dashed line is a scaled-down model of the
Galactic center density profile assuming gas feeding from massive
stars in which the bulk of the gas is thermally expelled in a wind
\citep{qua04}.  In this model the inner profile ($\lesssim 0.2$ pc) is
$\propto r^{-3/2}$, while the outer profile ($\gtrsim 0.4$ pc) has a
Wind ($\propto r^{-2}$) profile.  \label{fig:nr}} \end{figure}

In addition to the substantial increase in energy, we also find that
the normalization of the density profile, $n_{18}\equiv n_{\rm CNM}
r_{18}^2$, changes as a function of time (Figure~\ref{fig:params}).
This indicates that the radial density structure is not a simple Wind
profile, for which $n_{18}$ is by definition constant.  In particular,
we find that $n_{18}$ decreases at $5-10$ d, followed by a steady
increase at $10-216$ d is about a factor of 8.  The resulting radial
density profile is shown in Figure~\ref{fig:nr}, and clearly reflects
the complex evolution of $n_{18}$.  The density profile at $r\approx
0.1-0.4$ pc is $\rho\propto r^{-1.5}$, followed by a uniform density
at $r\approx 0.4-0.6$ pc, and a slow transition back to $\rho\propto
r^{-1.5}$ by $r\approx 1.2$ pc.  The inferred density is $n\approx
0.2-2$ cm$^{-3}$ at $r\approx 1.2-0.1$ pc, respectively.  These values
scale with the choice of equipartition fractions: for
$\epsilon_B=0.01$ the density is about 3 times larger, while for
$\epsilon_e=0.01$ it is about 1.7 times lower.  Thus, the inferred
density is fairly robust to large changes in the equipartition
fractions.

The flattening at $r\approx 0.4-0.6$ pc coincides with the rapid
increase in energy at $\delta t\approx 35-100$ d.  This can be
understood in the context of an outflow with a distribution of Lorentz
factors since the relative increase in density compared to the
previous $r^{-1.5}$ profile leads to enhanced contribution from lower
Lorentz factor ejecta.  A similar effect has been observed in the
radio emission from some core-collapse supernovae, which have a steep
ejecta profile with $E\propto v^{-5}$ (e.g.,
\citealt{che82,mm99,bkc02,sck+06}).

Also shown in Figure~\ref{fig:nr} are the inferred densities in the
central parsec of the Milky way galactic center from X-ray
measurements \citep{bmm+03}.  The Galactic Center density is similar
that inferred for \sw\ at $r\approx 0.05$ pc, but is about 30 times
larger at $r\approx 0.4$ pc.  If the bulk of the gas in the central
parsec is due to mass loss from massive stars (e.g.,
\citealt{mel92,bmm+03,qua04}), the lower density inferred here may
indicate a much smaller number of massive stars in the central parsec
of the host galaxy of \sw.  Although we cannot investigate star
formation on scales smaller than $\sim 1$ kpc in the host of \sw, the
overall star formation rate in this galaxy is indeed a factor of
several times lower than in the Milky Way \citep{ltc+11}.  Regardless
of the reason for the difference in density, it is quite remarkable
that the radio emission from \sw\ offers as detailed a view (or
better) of the density profile in the inner parsec around an inactive
SMBH at $z=0.354$ as available for the Galactic Center at a distance
of only 8.5 kpc.

\subsection{Optical and Near-Infrared Emission}
\label{sec:oir}

\begin{figure}
\epsscale{1}
\plotone{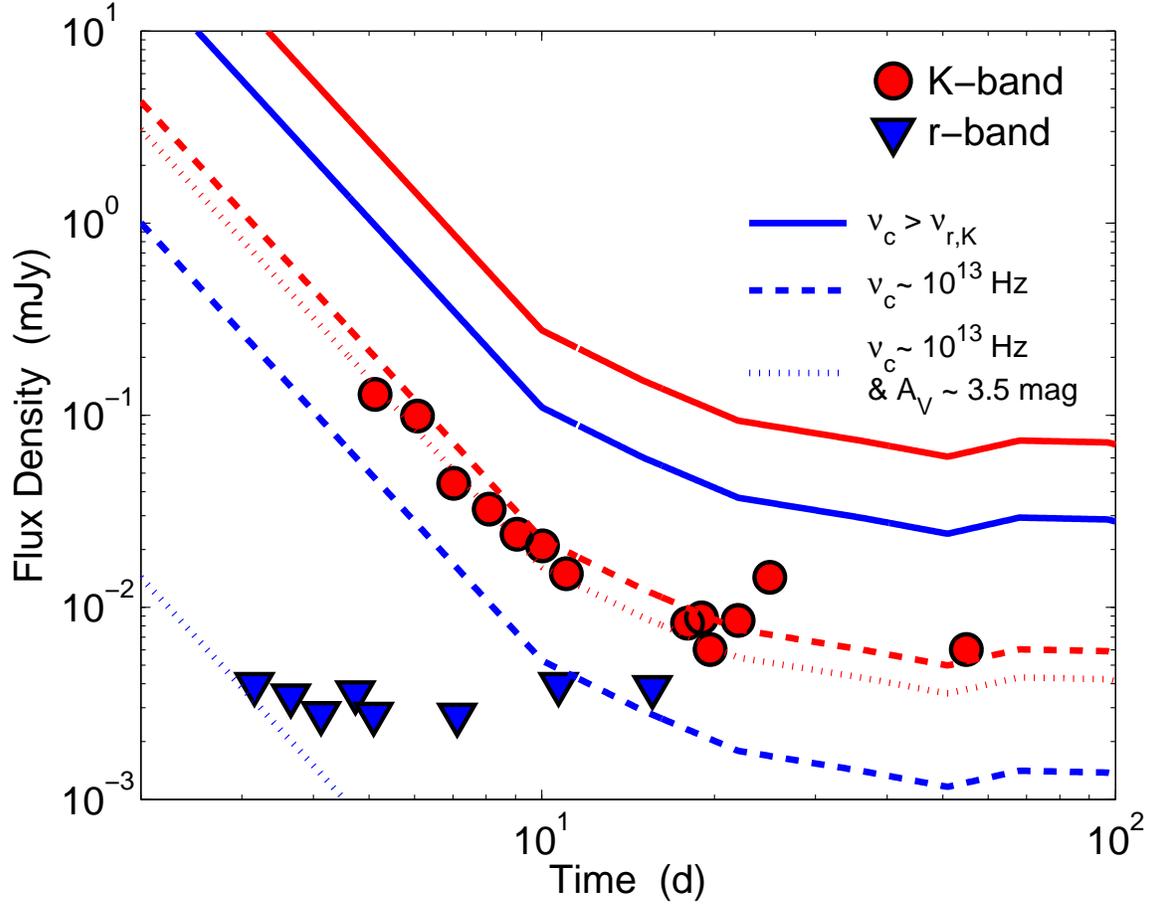}
\caption{Predicted optical ($r$-band; blue) and near-infrared
($K$-band; red) light curves using the results of the radio modeling.
The upper limits in $r$-band and detections in $K$-band are from
\citet{ltc+11}.  Since the $K$-band fluxes are the total for \sw\ and
its host galaxy we have subtracted an estimated host contribution of
about 20 $\mu$Jy ($K\approx 20.6$ AB mag).  The solid lines are models
without a cooling break between the radio and optical/near-IR, which
clearly over-estimate the $K$-band flux.  The dashed lines include a
cooling break at $\nu_c\approx 10^{13}$ Hz, and the dotted lines add
host galaxy extinction of $A_V\approx 3.5$ mag to account for the
optical non-detections.  The combination of a cooling break and
extinction provides an excellent fit to the near-IR evolution.
\label{fig:oir}} 
\end{figure}

Using the radio modeling we can also predict the optical and near-IR
emission from the jet.  In Figure~\ref{fig:oir} we plot the optical
$r$-band upper limits from \citet{ltc+11}, as well as their near-IR
$K$-band measurements.  The latter include the total flux from \sw\
and its host galaxy since transient-free templates are not presently
available.  We estimate the host contribution to be about $20$ $\mu$Jy
($K\approx 20.6$ AB mag) by requiring that the overall shape of the
$K$-band light curve match our predicted light curve.  We note that
this only affects the light curve shape at $\delta t\gtrsim 15$ d
since at earlier times the observed flux is dominated by the transient
itself.  A simple extrapolation of our model over-predicts the $K$-band
flux density by about an order of magnitude, indicating the presence
of an additional break in the spectrum between the radio and
optical/near-IR band.  Such a break is indicative of the synchrotron
cooling frequency, $\nu_c$.  To explain the $K$-band flux density
requires $\nu_c\sim 10^{13}$ Hz.  Even if we include this break, the
observed $r$-band limits are still a factor of about 70 times fainter
than the model prediction.  To explain this discrepancy requires host
galaxy extinction of $A_{\rm V,host}\gtrsim 3.5$ mag (see also
\citealt{ltc+11}, ZBS11).  With the addition of a cooling break and
host galaxy extinction we find an excellent match between our model
and the near-IR fluxes.  Since these data were not used in the
fitting, the agreement between our model and the data provides an
independent confirmation of the model.

\section{Implications for Relativistic Jets and the Environments of
Supermassive Black Holes} \label{sec:implic}

Our on-going radio observations have uncovered two unexpected and
critical results regarding the nature of the relativistic outflow from
\sw\ and the parsec-scale environment around a previously-dormant
SMBH.  We find that the relativistic outflow is not dominated by a
single Lorentz factor, as typically assumed in GRB jet models.
Instead, the outflow has a Lorentz factor distribution with
$E_j(>\Gamma_j)\propto\Gamma_j^{-2.5}$, at least over the range of
Lorentz factors probed so far, $\Gamma_j\approx 2.2-5.5$.  This
scenario is reminiscent of supernova ejecta, in which the energy is a
strong function of velocity, $E\propto v^{-5}$ \citep{che82,mm99}.  In
the SN case, this coupling reflects shock acceleration through the
steep density gradient in the outer envelope of the star.  In the case
of \sw\ we do not expect such a density gradient, suggesting instead
that the ejecta structure may be an intrinsic property of relativistic
jet launching, potentially through the Blandford-Znajek mechanism
\citep{bz77}.  In this case, the same structure may also apply to the
relativistic jets of GRBs, which are generally assumed to have a
single Lorentz factor.

To assess this possibility we compare the ratio of event duration to
the accretion disk dynamical timescale, $t_{\rm dyn}\propto M_{\rm
BH}(r/r_g)^{3/2}$, where $r_g=2GM_{\rm BH}/c^2$ is the black hole
Schwarzschild radius.  For similar ratios of $(r/r_g)$, since $M_{\rm
BH,Sw1644+57}\sim 10^6\,{\rm M}_\odot\sim 10^5M_{\rm BH,GRBs}$, the
dynamical timescale in \sw\ is about $10^5$ times longer than in GRBs.
The ratio of durations is similar, $\sim 10^6$ s for \sw\ and $\sim
10$ s for GRBs.  Therefore, the ratio of duration to dynamical
timescale is similar, and the jet launching mechanism may imprint a
similar profile in the case of \sw\ and GRBss.  Of course, in long
GRBs the jet still has to propagate through the stellar envelope,
which may modify the jet structure.

We next address the potential implications of the radial density
profile and the flattening at $r\approx 0.4-0.6$ pc.  One possibility
is that this scale represents the Bondi radius for spherical accretion
from the CNM, $r_B=GM/c_s^2=(c/c_s)^2\,r_g/2$; here $c_s$ is the sound
speed in the CNM and $r_g=2GM/c^2$ is the Schwarzschild radius of the
SMBH.  For a CNM gas temperature of $\sim 10^7$ K \citep{bmm+03,nf11},
we find $r_B\approx 10^6r_s\approx 0.1-1$ pc for the SMBH mass of
$10^6-10^7$ M$_\odot$ inferred for \sw.  Thus, the radius at which we
infer a density enhancement is in reasonable agreement with the
expected Bondi radius.  In addition, we note that the expected density
profile inside the Bondi radius, $\rho\propto r^{-3/2}$, matches the
inferred profile for \sw\ on a scale of $\sim 0.15-0.4$ pc.  In the
standard scenario, the medium outside the Bondi radius is assumed to
have a uniform density, while here we infer a continued decline in the
density at $r\approx 0.6-1.2$ pc.  This may reflect the complex
conditions in the inner few parsecs around the SMBH in which the
interstellar density profile may be mainly influenced by mass loss
from stars (as in the case of the Milky Way galactic center; e.g.,
\citealt{kgd+91}).  Indeed, from the inferred density of $n_{\rm
CNM}\approx 0.6$ cm$^{-3}$ at $r\approx 0.5$ pc, the required mass
loss rate is $\dot{M}\approx 5\times 10^{-5}$ M$_\odot$ yr$^{-1}$ for
a wind velocity of $10^3$ km s$^{-1}$.

In particular, in the case of the Galactic Center it has long been
believed that the gas in the central parsec is supplied by mass loss
from massive stars, with the bulk of the gas ($\gtrsim 90\%$) being
thermally expelled in a wind (\citealt{mel92,bmm+03,qua04}, but see
\citealt{ssr11} for an alternative interpretation of the X-ray
emission).  Using a spherically symmetric hydrodynamic simulation of
the gas supplied by stellar winds, \citet{qua04} showed that beyond
about 0.4 pc gas is mainly expelled with a resulting Wind profile
($\rho\propto r^{-2}$) on larger scales, while on scales of $\lesssim
0.2$ pc the profile is roughly as expected for Bondi accretion
($\rho\propto r^{-3/2}$); see Figure~\ref{fig:nr}.  The profile we
find here is somewhat different from this model, but this may be due
to a different distribution of massive stars, or to the simplifying
assumption of spherical symmetry in the model.

\section{Conclusions and Future Directions}
\label{sec:conc}

We presented radio observations of \sw\ extending to $\delta t\approx
216$ d and spanning a wide range of frequencies.  The evolution of the
radio emission changes dramatically at $\delta t \gtrsim 1$ month,
requiring an increase in the total energy by about an order of
magnitude, a density profile of $\rho\propto r^{-3/2}$ ($0.1-1.2$ pc),
and a flattening at $r\approx 0.4-0.6$ pc.  A comparison of the model
to optical limits and near-IR detections indicates a cooling break at
$\nu_c\sim 10^{13}$ Hz and host galaxy extinction of $A_{\rm V,host}
\gtrsim 3.5$ mag.

The increase in energy cannot be explained by injection from an
$L\propto t^{-5/3}$ tail that is expected in tidal disruption events
and which matches the evolution of the X-ray emission.  We conclude
that a natural explanation is a structured outflow with
$E(>\Gamma_j)\propto \Gamma_j^{-2.5}$.  The inferred density profile
and the radial scale of the density enhancement are in rough agreement
with the expectation for Bondi accretion from a circumnuclear medium.
The jet energetics and structure, as well as the detailed density
profile on $\sim 0.1-1$ pc scale are a testament to the important
insight that can be gained from continued radio observations of \sw.
In particular, the radial density profile is traced in greater detail
than even the inner parsec of the Milky Way.  Continued radio
observations will probe the environment to a scale of $\sim 10$ pc in
the coming decade.

\begin{figure}
\epsscale{1}
\plotone{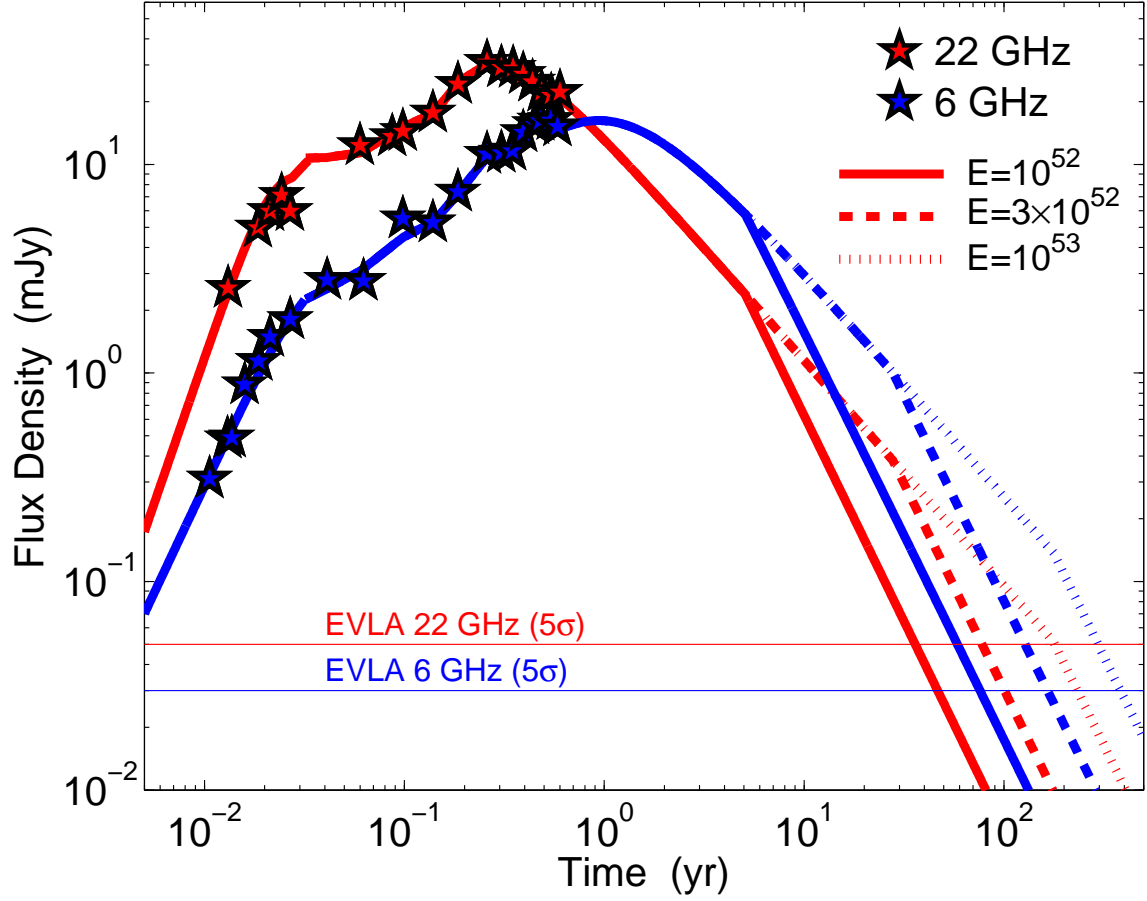}
\caption{Predicted evolution of the radio light curves at 6 GHz (blue)
and 22 GHz (red) assuming a radial density profile of $\rho\propto
r^{-1.5}$ at $r\gtrsim 1$ pc and three values for the maximum
integrated beaming-corrected energy (solid: $E_j=10^{52}$ erg; dashed:
$E_j=3\times 10^{52}$ erg; dotted: $E_j=10^{53}$ erg).  The thin
horizontal lines mark the $5\sigma$ sensitivity of the EVLA, and
indicate that the radio emission from \sw\ should be detectable for
decades (and perhaps centuries) at centimeter wavelengths.
\label{fig:predlc}} 
\end{figure}

Using the results of our analysis we can predict the future evolution
of the radio emission (modulo any future unpredictable changes in
energy and/or density as we have found here).  We use the evolution of
$L_{\rm j,iso}$ and $n_{\rm CNM}$ as inferred from the data at $\delta
t\lesssim 216$ d, and assume that the density will continue to evolve
as $\rho\propto r^{-1.5}$ and that the energy will increase to a
maximum beaming-corrected value of $E_j=L_{\rm j,iso}t_j[1-{\rm cos}
(\theta_j)]$ with $E_j=10^{52},\,3\times 10^{52},\,10^{53}$ erg.  The
resulting light curves at 6 and 22 GHz are shown in
Figure~\ref{fig:predlc}.  The long-term evolution is marked by a break
when $E_j$ achieves its maximum value, corresponding to about $5$,
$28$, and $180$ yr for our three choices of maximum energy.  Using the
$5\sigma$ sensitivity of the EVLA in an observation of a few
hours\footnotemark\footnotetext{At 22 GHz we use the sensitivity for
the full 8 GHz bandwidth that will become available some time in
2012.}, we find that the emission at 22 and 6 GHz should be detectable
for at least $\sim 40$ yr and $\sim 80$ yr, respectively.  Indeed, any
significant upgrades to the EVLA or the construction of more sensitive
radio facilities in the coming decades may extend the range of
detectability to centuries\footnotemark\footnotetext{Significant
budget cuts to radio facilities in the future may lead to the opposite
effect.}.  The same is true if the total energy scale is $\sim
10^{53}$ erg.

\begin{figure}
\epsscale{1}
\plotone{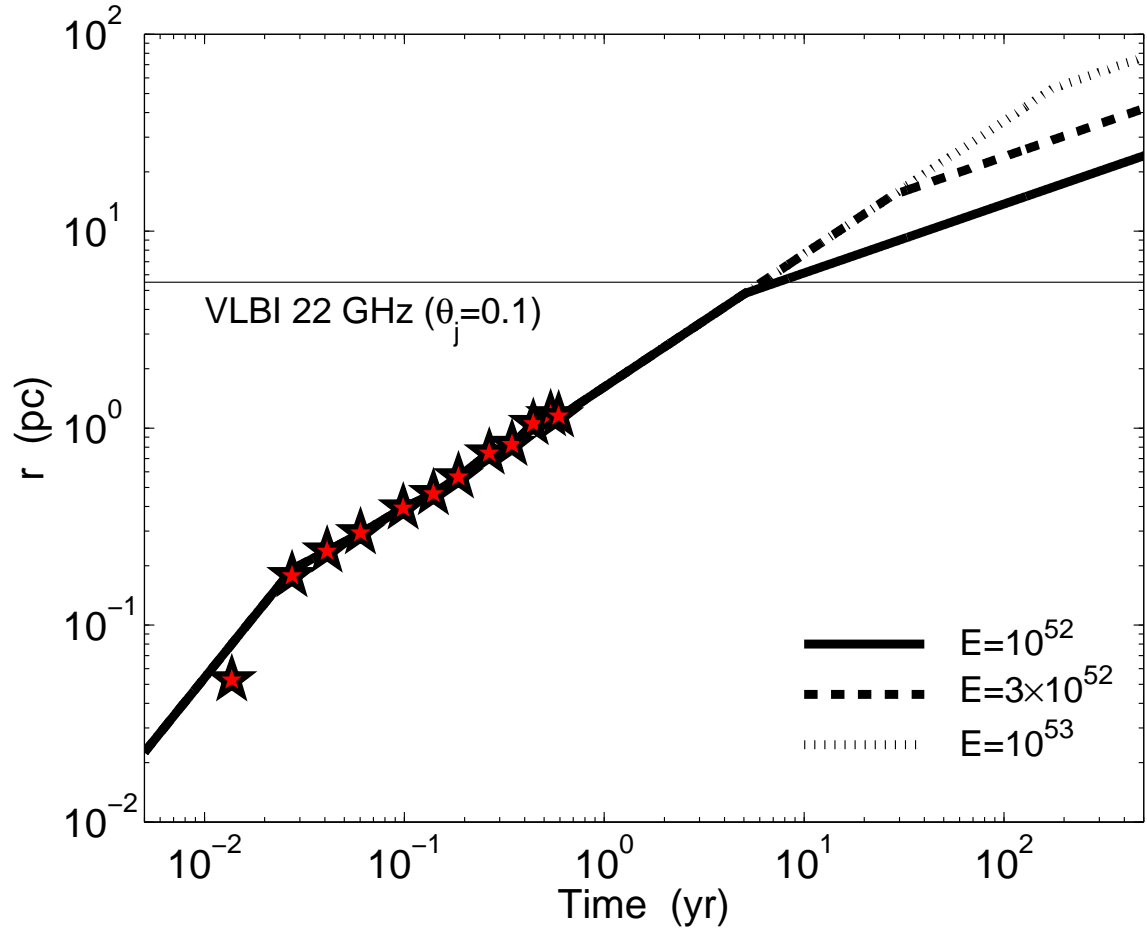}
\caption{Predicted evolution of the jet radius assuming a radial
density profile of $\rho\propto r^{-1.5}$ at $r\gtrsim 1$ pc and three
values for the maximum beaming energy (solid: $E_j=10^{52}$ erg;
dashed: $E_j=3\times 10^{52}$ erg; dotted: $E_j=10^{53}$ erg).  The
thin horizontal line marks the resolution of VLBI for a jet opening
angle of $\theta_j=0.1$.  The source should become resolvable at
$\delta t\sim 6$ yr, with an expected 22 GHz flux density of about 2
mJy (Figure~\ref{fig:predlc}).  If the jet instead begins to undergo
significant spreading it may become resolvable at $\sim 1$ yr when the
22 GHz flux density is still $\sim 10$ mJy.
\label{fig:predrad}} 
\end{figure}

An equally important question is whether the jet will be resolvable
with VLBI in the future.  The projected radius is $r_{\rm proj}\approx
r\theta_j$, as long as the jet maintains its collimation.  In
Figure~\ref{fig:predrad} we plot the predicted future evolution of $r$
using the prescription described above.  We find that for
$\theta_j=0.1$ and a best-case VLBI angular resolution of $\approx
0.2$ mas (FWHM), the source should become resolvable at $\delta
t\approx 6$ yr.  On this timescale the 22 GHz flux density is expected
to be only $\approx 2$ mJy (Figure~\ref{fig:predlc}), still accessible
with VLBI.  While the flux density at 6 GHz is expected to be larger
by about a factor of $2.6$, the angular resolution at this frequency
is poorer by about a factor of 3.7, making it less competitive than 22
GHz.  Thus, we conclude that the radio emission from \sw\ may be
marginally resolved in a few years.  On the other hand, if the jet
undergoes significant spreading on the timescale at which it becomes
non-relativistic (as expected for GRB jets: e.g., \citealt{lw00}) it
is possible that it will become resolvable at $\delta t\sim 1-2$ yr
when the expected 22 GHz flux density is still $\sim 10$ mJy.

We are undertaking continued multi-frequency radio monitoring of \sw\
to follow the long-term evolution of the relativistic outflow and the
radial profile of the ambient medium.  Even in the absence of any
future dramatic changes relative to the current evolution, we expect
that in the next few years we may be able to determine the total
energy of the relativistic outflow, measure the spreading of the jet,
and study the radial density profile to a scale of $\sim 10$ pc.
Future papers in this series will detail these results.

\acknowledgments We thank Ramesh Narayan, Ryan Chornock, and Nicholas
Stone for helpful discussions, and Glen Petitpas for assistance with
the SMA data reduction.  E.B.~acknowledges support from Swift AO6
grant NNX10AI24G and from the National Science Foundation through
Grant AST-1107973.  A.B. is supported by a Marie Curie Outgoing
International Fellowship (FP7) of the European Union (project number
275596).  This work is partially based on observations with the 100-m
telescope of the MPIfR (Max-Planck-Institut f$\ddot{\rm u}$r
Radioastronomie) at Effelsberg.  The Submillimeter Array is a joint
project between the Smithsonian Astrophysical Observatory and the
Academia Sinica Institute of Astronomy and Astrophysics, and is funded
by the Smithsonian Institution and the Academia Sinica.  Support for
CARMA construction was derived from the Gordon and Betty Moore
Foundation, the Kenneth T.~and Eileen L.~Norris Foundation, the James
S.~McDonnell Foundation, the Associates of the California Institute of
Technology, the University of Chicago, the states of California,
Illinois, and Maryland, and the National Science Foundation. Ongoing
CARMA development and operations are supported by the National Science
Foundation under a cooperative agreement, and by the CARMA partner
universities.  The AMI arrays are supported by the University of
Cambridge and the STFC.  This work made use of data supplied by the UK
Swift Science Data Centre at the University of Leicester.


\clearpage
\LongTables
\begin{deluxetable}{rccr}
\tabletypesize{\footnotesize}
\tablecolumns{4} 
\tabcolsep0.15in\footnotesize
\tablewidth{0pt} 
\tablecaption{Radio Observations of \sw\
\label{tab:data}}
\tablehead{
\colhead{$\delta t\,^a$} &
\colhead{Facility}       &
\colhead{Frequency}      &
\colhead{Flux Density}   \\
\colhead{(d)}            &              
\colhead{}               &            
\colhead{(GHz)}          &            
\colhead{(mJy)}            
}
\startdata
    6.79 & EVLA & 1.4 & $0.21\pm 0.08$ \\
  126.59 & EVLA & 1.4 & $1.10\pm 0.10$ \\
  174.47 & EVLA & 1.4 & $1.60\pm 0.11$ \\
  197.41 & EVLA & 1.4 & $1.80\pm 0.10$ \\\hline\hline
    3.87 & EVLA & 4.9 & $ 0.25\pm 0.02$ \\
    4.76 & EVLA & 4.9 & $ 0.34\pm 0.02$ \\
    5.00 & EVLA & 4.9 & $ 0.34\pm 0.02$ \\
    5.79 & EVLA & 4.9 & $ 0.61\pm 0.02$ \\
    6.78 & EVLA & 4.9 & $ 0.82\pm 0.02$ \\
    7.77 & EVLA & 4.9 & $ 1.48\pm 0.02$ \\
    9.79 & EVLA & 4.9 & $ 1.47\pm 0.02$ \\
   14.98 & EVLA & 4.9 & $ 1.80\pm 0.03$ \\
   22.78 & EVLA & 4.9 & $ 2.10\pm 0.01$ \\
   35.86 & EVLA & 4.9 & $ 4.62\pm 0.02$ \\
   50.65 & EVLA & 4.9 & $ 4.84\pm 0.03$ \\
   67.61 & EVLA & 4.9 & $ 5.86\pm 0.03$ \\
   94.64 & EVLA & 4.9 & $ 9.06\pm 0.03$ \\
  111.62 & EVLA & 4.9 & $ 9.10\pm 0.03$ \\
  126.51 & EVLA & 4.9 & $ 9.10\pm 0.03$ \\
  143.62 & EVLA & 4.9 & $11.71\pm 0.03$ \\
  164.38 & EVLA & 4.9 & $12.93\pm 0.05$ \\
  174.47 & EVLA & 4.9 & $12.83\pm 0.06$ \\
  197.41 & EVLA & 4.9 & $13.29\pm 0.03$ \\
  213.32 & EVLA & 4.9 & $12.43\pm 0.04$ \\\hline\hline
    3.87 & EVLA & 6.7 & $ 0.38\pm 0.02$ \\
    4.76 & EVLA & 6.7 & $ 0.63\pm 0.02$ \\
    5.00 & EVLA & 6.7 & $ 0.64\pm 0.02$ \\
    5.79 & EVLA & 6.7 & $ 1.16\pm 0.02$ \\
    6.79 & EVLA & 6.7 & $ 1.47\pm 0.02$ \\
    7.77 & EVLA & 6.7 & $ 1.50\pm 0.02$ \\
    9.79 & EVLA & 6.7 & $ 2.15\pm 0.02$ \\
   14.98 & EVLA & 6.7 & $ 3.79\pm 0.03$ \\
   22.78 & EVLA & 6.7 & $ 3.44\pm 0.01$ \\
   35.86 & EVLA & 6.7 & $ 6.39\pm 0.02$ \\
   50.65 & EVLA & 6.7 & $ 5.70\pm 0.02$ \\
   67.61 & EVLA & 6.7 & $ 8.94\pm 0.03$ \\
   94.64 & EVLA & 6.7 & $13.43\pm 0.03$ \\
  111.62 & EVLA & 6.7 & $13.66\pm 0.03$ \\
  126.51 & EVLA & 6.7 & $14.16\pm 0.04$ \\
  143.62 & EVLA & 6.7 & $16.85\pm 0.04$ \\
  164.38 & EVLA & 6.7 & $18.27\pm 0.06$ \\
  174.47 & EVLA & 6.7 & $19.59\pm 0.16$ \\
  197.41 & EVLA & 6.7 & $19.34\pm 0.03$ \\
  213.32 & EVLA & 6.7 & $18.02\pm 0.05$ \\\hline\hline
   14.97 & EVLA & 8.4 & $ 5.49\pm 0.09$ \\
  127.69 & EVLA & 8.4 & $19.03\pm 0.14$ \\
  159.77 & EVLA & 8.4 & $22.15\pm 0.20$ \\
  174.47 & EVLA & 8.4 & $23.19\pm 0.38$ \\
  177.50 & EVLA & 8.4 & $23.65\pm 0.16$ \\
  197.41 & EVLA & 8.4 & $22.42\pm 0.10$ \\
  213.32 & EVLA & 8.4 & $22.04\pm 0.13$ \\
  219.22 & EVLA & 8.4 & $21.52\pm 0.09$ \\\hline\hline
    5.81 & AMI-LA & 15.4 & $ 2.69\pm 0.44$ \\
    6.64 & AMI-LA & 15.4 & $ 3.62\pm 0.22$ \\
    7.62 & AMI-LA & 15.4 & $ 4.32\pm 0.28$ \\
    8.55 & AMI-LA & 15.4 & $ 5.07\pm 0.36$ \\
    9.56 & AMI-LA & 15.4 & $ 6.68\pm 0.51$ \\
   10.78 & AMI-LA & 15.4 & $ 6.74\pm 0.48$ \\
   11.56 & AMI-LA & 15.4 & $ 7.50\pm 0.44$ \\
   13.64 & AMI-LA & 15.4 & $ 8.02\pm 0.71$ \\
   14.57 & AMI-LA & 15.4 & $ 8.43\pm 0.32$ \\
   16.47 & AMI-LA & 15.4 & $ 8.86\pm 0.58$ \\
   18.65 & AMI-LA & 15.4 & $ 8.62\pm 0.40$ \\
   19.73 & AMI-LA & 15.4 & $10.04\pm 0.63$ \\
   21.78 & AMI-LA & 15.4 & $10.91\pm 0.89$ \\
   22.76 & AMI-LA & 15.4 & $11.01\pm 0.77$ \\
   25.38 & AMI-LA & 15.4 & $10.28\pm 0.58$ \\
   26.74 & AMI-LA & 15.4 & $11.36\pm 0.80$ \\
   31.45 & AMI-LA & 15.4 & $11.24\pm 0.41$ \\
   33.61 & AMI-LA & 15.4 & $12.14\pm 0.16$ \\
   34.79 & AMI-LA & 15.4 & $11.89\pm 0.25$ \\
   35.71 & AMI-LA & 15.4 & $13.39\pm 0.49$ \\
   37.49 & AMI-LA & 15.4 & $13.63\pm 0.45$ \\
   38.60 & AMI-LA & 15.4 & $13.72\pm 0.39$ \\
   39.78 & AMI-LA & 15.4 & $10.80\pm 0.91$ \\
   40.67 & AMI-LA & 15.4 & $13.38\pm 0.40$ \\
   41.69 & AMI-LA & 15.4 & $13.64\pm 0.21$ \\
   43.63 & AMI-LA & 15.4 & $14.21\pm 1.24$ \\
   45.74 & AMI-LA & 15.4 & $12.33\pm 0.37$ \\
   48.69 & AMI-LA & 15.4 & $14.06\pm 0.41$ \\
   49.64 & AMI-LA & 15.4 & $13.94\pm 0.05$ \\
   50.54 & AMI-LA & 15.4 & $14.39\pm 0.25$ \\
   53.39 & AMI-LA & 15.4 & $15.94\pm 0.59$ \\
   55.62 & AMI-LA & 15.4 & $15.28\pm 0.33$ \\
   56.52 & AMI-LA & 15.4 & $17.91\pm 0.07$ \\
   60.73 & AMI-LA & 15.4 & $15.00\pm 1.18$ \\
   62.54 & AMI-LA & 15.4 & $19.23\pm 0.62$ \\
   63.45 & AMI-LA & 15.4 & $16.47\pm 0.21$ \\
   65.51 & AMI-LA & 15.4 & $18.77\pm 0.37$ \\
   68.60 & AMI-LA & 15.4 & $19.06\pm 0.48$ \\
   70.69 & AMI-LA & 15.4 & $20.34\pm 0.38$ \\
   71.38 & AMI-LA & 15.4 & $19.36\pm 0.56$ \\
   73.48 & AMI-LA & 15.4 & $21.27\pm 0.34$ \\
   74.48 & AMI-LA & 15.4 & $21.84\pm 0.59$ \\
   76.65 & AMI-LA & 15.4 & $21.40\pm 0.51$ \\
   77.47 & AMI-LA & 15.4 & $23.08\pm 0.26$ \\
   78.49 & AMI-LA & 15.4 & $23.26\pm 0.65$ \\
   79.49 & AMI-LA & 15.4 & $23.22\pm 0.24$ \\
   80.47 & AMI-LA & 15.4 & $22.94\pm 0.50$ \\
   81.48 & AMI-LA & 15.4 & $21.74\pm 0.45$ \\
   83.42 & AMI-LA & 15.4 & $23.99\pm 0.50$ \\
   86.65 & AMI-LA & 15.4 & $22.94\pm 0.33$ \\
   88.37 & AMI-LA & 15.4 & $24.69\pm 0.26$ \\
   89.64 & AMI-LA & 15.4 & $25.90\pm 0.50$ \\
   91.58 & AMI-LA & 15.4 & $26.07\pm 1.04$ \\
   92.58 & AMI-LA & 15.4 & $25.40\pm 0.55$ \\
   95.39 & AMI-LA & 15.4 & $25.34\pm 0.72$ \\
   98.42 & AMI-LA & 15.4 & $26.15\pm 0.25$ \\
  100.61 & AMI-LA & 15.4 & $27.83\pm 0.57$ \\
  101.60 & AMI-LA & 15.4 & $25.96\pm 0.91$ \\
  102.60 & AMI-LA & 15.4 & $26.60\pm 0.26$ \\
  105.59 & AMI-LA & 15.4 & $26.68\pm 0.21$ \\
  107.38 & AMI-LA & 15.4 & $27.89\pm 0.33$ \\
  108.32 & AMI-LA & 15.4 & $26.84\pm 0.40$ \\
  110.34 & AMI-LA & 15.4 & $28.24\pm 0.22$ \\
  112.50 & AMI-LA & 15.4 & $27.10\pm 0.58$ \\
  113.50 & AMI-LA & 15.4 & $28.82\pm 0.31$ \\
  115.49 & AMI-LA & 15.4 & $28.39\pm 0.23$ \\
  118.31 & AMI-LA & 15.4 & $28.58\pm 0.47$ \\
  119.48 & AMI-LA & 15.4 & $26.90\pm 0.80$ \\
  120.55 & AMI-LA & 15.4 & $27.88\pm 0.35$ \\
  121.55 & AMI-LA & 15.4 & $26.92\pm 0.73$ \\
  122.18 & AMI-LA & 15.4 & $29.87\pm 0.61$ \\
  123.45 & AMI-LA & 15.4 & $28.86\pm 0.41$ \\
  124.38 & AMI-LA & 15.4 & $27.46\pm 0.97$ \\
  125.51 & AMI-LA & 15.4 & $27.40\pm 0.56$ \\
  127.38 & AMI-LA & 15.4 & $28.96\pm 0.39$ \\
  128.35 & AMI-LA & 15.4 & $28.69\pm 1.00$ \\
  130.38 & AMI-LA & 15.4 & $27.87\pm 0.64$ \\
  131.41 & AMI-LA & 15.4 & $28.94\pm 1.11$ \\
  132.49 & AMI-LA & 15.4 & $29.39\pm 0.96$ \\
  133.35 & AMI-LA & 15.4 & $30.81\pm 1.03$ \\
  134.36 & AMI-LA & 15.4 & $29.73\pm 0.52$ \\
  135.32 & AMI-LA & 15.4 & $31.25\pm 0.59$ \\
  136.46 & AMI-LA & 15.4 & $29.31\pm 1.29$ \\
  137.51 & AMI-LA & 15.4 & $29.58\pm 0.23$ \\
  138.51 & AMI-LA & 15.4 & $28.50\pm 0.69$ \\
  139.51 & AMI-LA & 15.4 & $28.96\pm 0.57$ \\
  140.50 & AMI-LA & 15.4 & $29.22\pm 0.52$ \\
  141.13 & AMI-LA & 15.4 & $29.03\pm 0.45$ \\
  142.30 & AMI-LA & 15.4 & $26.60\pm 0.31$ \\
  143.26 & AMI-LA & 15.4 & $28.96\pm 0.61$ \\
  144.42 & AMI-LA & 15.4 & $28.36\pm 0.57$ \\
  146.48 & AMI-LA & 15.4 & $28.92\pm 0.64$ \\
  147.48 & AMI-LA & 15.4 & $27.44\pm 0.76$ \\
  148.27 & AMI-LA & 15.4 & $29.49\pm 0.97$ \\
  149.27 & AMI-LA & 15.4 & $29.82\pm 0.43$ \\
  150.47 & AMI-LA & 15.4 & $29.83\pm 0.23$ \\
  151.22 & AMI-LA & 15.4 & $29.31\pm 0.73$ \\
  152.27 & AMI-LA & 15.4 & $28.10\pm 1.30$ \\
  153.46 & AMI-LA & 15.4 & $26.71\pm 0.38$ \\
  155.45 & AMI-LA & 15.4 & $27.80\pm 0.45$ \\
  156.35 & AMI-LA & 15.4 & $29.92\pm 0.89$ \\
  157.45 & AMI-LA & 15.4 & $27.77\pm 0.31$ \\
  158.29 & AMI-LA & 15.4 & $27.72\pm 0.83$ \\
  159.28 & AMI-LA & 15.4 & $27.98\pm 0.59$ \\
  160.27 & AMI-LA & 15.4 & $28.93\pm 0.44$ \\
  161.20 & AMI-LA & 15.4 & $26.94\pm 0.84$ \\
  162.21 & AMI-LA & 15.4 & $26.81\pm 1.21$ \\
  163.28 & AMI-LA & 15.4 & $28.82\pm 0.68$ \\
  166.00 & AMI-LA & 15.4 & $26.92\pm 0.58$ \\
  168.13 & AMI-LA & 15.4 & $25.90\pm 0.64$ \\
  169.13 & AMI-LA & 15.4 & $25.47\pm 0.83$ \\
  172.25 & AMI-LA & 15.4 & $26.12\pm 0.48$ \\
  174.33 & AMI-LA & 15.4 & $26.12\pm 0.15$ \\
  177.32 & AMI-LA & 15.4 & $26.31\pm 0.37$ \\
  181.30 & AMI-LA & 15.4 & $24.37\pm 0.85$ \\
  184.03 & AMI-LA & 15.4 & $23.06\pm 0.77$ \\
  187.19 & AMI-LA & 15.4 & $24.24\pm 0.43$ \\
  187.91 & AMI-LA & 15.4 & $24.28\pm 0.45$ \\
  189.37 & AMI-LA & 15.4 & $24.60\pm 0.59$ \\
  193.19 & AMI-LA & 15.4 & $26.15\pm 0.75$ \\
  194.01 & AMI-LA & 15.4 & $24.53\pm 0.34$ \\
  197.02 & AMI-LA & 15.4 & $26.72\pm 0.52$ \\
  200.01 & AMI-LA & 15.4 & $25.90\pm 0.15$ \\
  203.01 & AMI-LA & 15.4 & $24.12\pm 0.41$ \\
  207.08 & AMI-LA & 15.4 & $25.06\pm 0.45$ \\
  210.19 & AMI-LA & 15.4 & $24.24\pm 0.31$ \\
  214.17 & AMI-LA & 15.4 & $22.64\pm 0.38$ \\
  217.16 & AMI-LA & 15.4 & $24.16\pm 0.25$ \\
  220.20 & AMI-LA & 15.4 & $23.74\pm 0.05$ \\
  221.22 & AMI-LA & 15.4 & $24.33\pm 0.49$ \\
  225.91 & AMI-LA & 15.4 & $23.80\pm 0.14$ \\
  232.87 & AMI-LA & 15.4 & $23.72\pm 0.16$ \\
  235.01 & AMI-LA & 15.4 & $21.64\pm 0.16$ \\
  237.85 & AMI-LA & 15.4 & $20.86\pm 0.75$ \\\hline\hline
    4.79 & EVLA & 19.1 & $ 2.12\pm 0.02$ \\
    6.75 & EVLA & 19.1 & $ 4.36\pm 0.05$ \\
    7.77 & EVLA & 19.1 & $ 5.25\pm 0.03$ \\
    8.87 & EVLA & 19.1 & $ 6.38\pm 0.06$ \\
    9.78 & EVLA & 19.1 & $ 5.42\pm 0.03$ \\
   21.89 & EVLA & 19.1 & $12.01\pm 0.03$ \\
   31.74 & EVLA & 19.1 & $13.50\pm 0.05$ \\
   35.86 & EVLA & 19.1 & $13.97\pm 0.05$ \\
   50.65 & EVLA & 19.1 & $17.11\pm 0.06$ \\
   67.61 & EVLA & 19.1 & $23.03\pm 0.06$ \\
   94.64 & EVLA & 19.1 & $31.36\pm 0.07$ \\
  111.62 & EVLA & 19.1 & $30.21\pm 0.10$ \\
  127.83 & EVLA & 19.1 & $29.75\pm 0.22$ \\
  142.62 & EVLA & 19.1 & $29.57\pm 0.13$ \\
  159.77 & EVLA & 19.1 & $26.10\pm 0.26$ \\
  177.50 & EVLA & 19.1 & $24.24\pm 0.16$ \\
  198.22 & EVLA & 19.1 & $23.02\pm 0.12$ \\
  219.22 & EVLA & 19.1 & $23.15\pm 0.07$ \\\hline\hline
    4.79 & EVLA & 24.4 & $ 3.01\pm 0.03$ \\
    6.75 & EVLA & 24.4 & $ 5.58\pm 0.06$ \\
    7.77 & EVLA & 24.4 & $ 6.70\pm 0.03$ \\
    8.87 & EVLA & 24.4 & $ 7.88\pm 0.12$ \\
    9.78 & EVLA & 24.4 & $ 6.84\pm 0.03$ \\
   21.89 & EVLA & 24.4 & $12.69\pm 0.02$ \\
   31.74 & EVLA & 24.4 & $13.80\pm 0.06$ \\
   35.86 & EVLA & 24.4 & $14.95\pm 0.05$ \\
   50.65 & EVLA & 24.4 & $18.30\pm 0.06$ \\
   67.61 & EVLA & 24.4 & $25.62\pm 0.06$ \\
   94.64 & EVLA & 24.4 & $30.67\pm 0.07$ \\
  111.62 & EVLA & 24.4 & $28.20\pm 0.13$ \\
  127.83 & EVLA & 24.4 & $28.29\pm 0.33$ \\
  142.62 & EVLA & 24.4 & $24.73\pm 0.16$ \\
  159.77 & EVLA & 24.4 & $23.83\pm 0.40$ \\
  177.50 & EVLA & 24.4 & $20.40\pm 0.19$ \\
  198.22 & EVLA & 24.4 & $19.88\pm 0.15$ \\
  219.22 & EVLA & 24.4 & $21.40\pm 0.08$ \\\hline\hline
    5.75 & EVLA & 43.6 & $ 8.11\pm 0.16$ \\
    6.75 & EVLA & 43.6 & $ 7.70\pm 0.14$ \\
    8.87 & EVLA & 43.6 & $ 9.62\pm 0.14$ \\
  111.62 & EVLA & 43.6 & $22.09\pm 0.65$ \\
  127.83 & EVLA & 43.6 & $20.84\pm 0.48$ \\
  159.77 & EVLA & 43.6 & $15.43\pm 0.63$ \\
  177.50 & EVLA & 43.6 & $15.73\pm 0.38$ \\
  198.22 & EVLA & 43.6 & $11.70\pm 0.26$ \\
  219.22 & EVLA & 43.6 & $15.32\pm 0.15$ \\\hline\hline
    4.90 & CARMA & 87 & $15.66\pm 0.51$ \\
    8.19 & CARMA & 87 & $20.16\pm 0.96$ \\
    9.14 & CARMA & 87 & $21.67\pm 0.33$ \\
   10.23 & CARMA & 87 & $14.70\pm 1.28$ \\
   12.14 & CARMA & 87 & $17.67\pm 0.94$ \\
   17.66 & CARMA & 87 & $12.15\pm 1.32$ \\
   22.11 & CARMA & 87 & $11.70\pm 1.74$ \\
   25.12 & CARMA & 87 & $16.95\pm 3.48$ \\
   99.75 & CARMA & 87 & $18.79\pm 0.65$ \\
  131.52 & CARMA & 87 & $15.18\pm 0.35$ \\
  148.66 & CARMA & 87 & $11.88\pm 0.35$ \\
  175.56 & CARMA & 87 & $ 9.39\pm 0.31$ \\\hline\hline
   19.25 & SMA & 200 & $14.10\pm 1.50$ \\
   24.32 & SMA & 200 & $10.70\pm 1.00$ \\\hline\hline
   10.30 & SMA & 230 & $14.90\pm 1.50$ \\
   11.13 & SMA & 230 & $11.70\pm 1.40$ \\
   17.23 & SMA & 230 & $13.30\pm 1.50$ \\
   18.25 & SMA & 230 & $ 9.90\pm 1.40$ \\
   20.24 & SMA & 230 & $ 8.20\pm 1.40$ \\
   21.25 & SMA & 230 & $ 8.30\pm 2.20$ \\
  125.05 & SMA & 230 & $ 6.10\pm 0.65$ \\\hline\hline
    5.13 & SMA & 345 & $35.10\pm 0.80$
\enddata
\tablecomments{$^a$ All values of $\delta t$ are relative to the
initial $\gamma$-ray detection: 2011 March 25.5 UT.}

\end{deluxetable}

\clearpage
\begin{deluxetable}{ccccccccccc}
\tabletypesize{\footnotesize}
\tablecolumns{11} 
\tabcolsep0.1in\footnotesize
\tablewidth{0pt} 
\tablecaption{Results of Broad-band Spectral Energy Distribution Fits
\label{tab:params}}
\tablehead{
\colhead{$\delta t$}                     &
\colhead{${\rm log}(\nu_a)$}             &
\colhead{${\rm log}(\nu_m)$}             &
\colhead{${\rm log}(F_{\nu_a})$}         &       
\colhead{${\rm log}(r_{18})$}            &      
\colhead{${\rm log}(\Gamma_{\rm sh})$}   &      
\colhead{${\rm log}(\Gamma_j)$}          &      
\colhead{${\rm log}(L_{\rm j,iso,48})$}  &      
\colhead{${\rm log}(n_{18})$}            &      
\colhead{${\rm log}(n_{\rm CNM})$}       &      
\colhead{${\rm log}(n_j)$}               \\      
\colhead{(d)}                            &              
\colhead{(Hz)}                           &            
\colhead{(Hz)}                           &            
\colhead{(mJy)}                          &           
\colhead{(cm)}                           &                    
\colhead{}                               &   
\colhead{}                               &   
\colhead{(erg s$^{-1}$)}                 &         
\colhead{(cm$^{-3}$)}                    &           
\colhead{(cm$^{-3}$)}                    &         
\colhead{(cm$^{-3}$)}     
}
\startdata
$  5$ & $11.01$ & $11.74$ & $1.47$ & $-0.79$ & $0.65$ & $0.78$ & $-1.33$ & $ 0.28$ & $ 1.82$ & $3.53$ \\
$ 10$ & $10.11$ & $11.41$ & $0.96$ & $-0.26$ & $0.61$ & $0.74$ & $-1.39$ & $-0.28$ & $ 0.25$ & $2.39$ \\
$ 15$ & $10.02$ & $11.21$ & $0.98$ & $-0.13$ & $0.55$ & $0.65$ & $-1.33$ & $-0.22$ & $ 0.05$ & $2.19$ \\
$ 22$ & $ 9.96$ & $10.99$ & $0.97$ & $-0.04$ & $0.51$ & $0.60$ & $-1.25$ & $-0.15$ & $-0.07$ & $2.08$ \\
$ 36$ & $ 9.95$ & $10.78$ & $1.11$ & $ 0.08$ & $0.47$ & $0.54$ & $-1.05$ & $ 0.01$ & $-0.17$ & $2.03$ \\
$ 51$ & $ 9.96$ & $10.62$ & $1.21$ & $ 0.16$ & $0.43$ & $0.50$ & $-0.92$ & $ 0.13$ & $-0.18$ & $2.01$ \\
$ 68$ & $ 9.99$ & $10.53$ & $1.39$ & $ 0.24$ & $0.41$ & $0.47$ & $-0.72$ & $ 0.28$ & $-0.20$ & $2.04$ \\
$ 97$ & $ 9.95$ & $10.39$ & $1.52$ & $ 0.36$ & $0.40$ & $0.45$ & $-0.53$ & $ 0.39$ & $-0.34$ & $1.99$ \\
$126$ & $ 9.97$ & $10.26$ & $1.58$ & $ 0.41$ & $0.36$ & $0.41$ & $-0.46$ & $ 0.49$ & $-0.33$ & $1.97$ \\
$161$ & $ 9.82$ & $10.13$ & $1.51$ & $ 0.52$ & $0.36$ & $0.40$ & $-0.39$ & $ 0.44$ & $-0.60$ & $1.82$ \\
$197$ & $ 9.83$ & $10.04$ & $1.56$ & $ 0.56$ & $0.34$ & $0.38$ & $-0.32$ & $ 0.52$ & $-0.60$ & $1.81$ \\
$216$ & $ 9.90$ & $ 9.99$ & $1.63$ & $ 0.55$ & $0.32$ & $0.35$ & $-0.29$ & $ 0.60$ & $-0.50$ & $1.85$ 
\enddata
\tablecomments{Inferred parameters of the relativistic outflow and
environment of \sw\ from model fits of individual multi-frequency
epochs.  The model is described in \S\ref{sec:model}.}
\end{deluxetable}

\clearpage

\clearpage

\clearpage

\clearpage

\clearpage

\clearpage

\clearpage

\clearpage

\clearpage

\end{document}